\documentclass[aps,pra,twocolumn,showpacs,preprintnumbers,runinaddress,lengthcheck]{revtex4}

\usepackage{amsmath}
\usepackage{graphicx}
\usepackage{dcolumn}
\usepackage{bm}
\usepackage{bbm}
\usepackage{appendix}

\begin{document}

\title{Security of Binary Modulated Continuous Variable Quantum Key Distribution under Collective Attacks}

\author{Yi-Bo \surname{Zhao}$^1$}
\author{Matthias \surname{Heid}$^{2,3}$}
\author{Johannes \surname{Rigas}$^4$}
\author{Norbert \surname{L\"utkenhaus}$^{2,3}$}

\affiliation{$^1$Key Lab of Quantum Information, University of Science and Technology of China, (CAS), Hefei, Anhui 230026, China\\
$^2$ Quantum Information Theory Group, Institut f\"{u}r Theoretische Physik I, \& Max-Planck Research Group, Institute of Optics, Information and Photonics, Universit\"{a}t Erlangen-N\"{u}rnberg, Staudtstra{\ss}e 7/B2, 91058 Erlangen, Germany \\
$^3$ Institute for Quantum Computing \& Department of Physics and Astronomy, University of Waterloo, 200 University Ave.~W.~N2L 3G1, Canada\\
$^4$ Departamento de \'Optica, Facultad de F\'isica, Universidad Complutense, 28040 Madrid, Spain}

\begin{abstract}
We give an achievable secret key rate of a binary modulated continuous variable quantum key distribution schemes in the collective attack scenario considering quantum channels that impose arbitrary noise on the exchanged signals. Bob performs homodyne measurements on the received states and the two honest parties employ a reverse reconciliation procedure in the classical post-processing step of the protocol.
\end{abstract}

\maketitle

\section{Introduction}

Quantum key distribution (QKD) is a way to establish a key between two communicating parties, traditionally called Alice and Bob, which is provable secure against any eavesdropping strategy of an technologically unlimited third party Eve. In principle, Alice and Bob can achieve this goal by exchanging nonorthogonal quantum states as signals and using non-commuting measurements on the receiver side. Any eavesdropper needs to interact with these quantum signals to gain information about the sent signal. This inevitably causes a disturbance of the signals and leads to errors in the data that Alice and Bob observe. If the amount of errors lies below a certain threshold, Alice and Bob proceed by post-processing their data: they correct for errors and employ privacy amplification to cut out any residual information that Eve might have with the raw key.
In this article, we give a lower bound to the secret key rate of a continuous variable (CV) QKD scheme \cite{assche05a,grosshans05a,heid06a,heid07a} employing homodyne detection in the collective attack scenario. As the outcomes of Bob's measurement are continuous, it is convenient to characterize Eve's interference with the signal states by the first and second moments of Bob's measurement outcomes. These parameters are usually given in terms of the observed loss and the excess noise of the quantum channel connecting Alice and Bob. The proof technique presented here can be used to compute secret key rates of a binary modulated CV-QKD scheme for arbitrary, in particular non-Gaussian observations, thereby extending the results given in \cite{heid06a}. This is important from a conceptual point of view, as the optimality of Gaussian attacks  \cite{garcia06a,navascues06a} has only been shown for CV-schemes using a Gaussian modulated set of coherent states as input \cite{grosshans03a,lance05a,heid07a}. So far, the security of the binary scheme is not fully established yet, even if one limits the eavesdropper to collective attacks. Our analysis presented here is restricted to the asymptotic key limit as the number of exchanged signals $n$ approaches infinity.

We consider the class of collective attacks \cite{renner05b,devetak05a}, thereby limiting Eve's possible interaction with the signal states. In this scenario, Eve can only interact with each signal individually, but she can store these quantum states for later usage. In the classical post-processing phase, Alice and Bob exchange information about their shared bit strings over a authenticated  classical channel. This information eventually leaks to Eve, who can make use of this additional knowledge to employ optimized measurement on her quantum states. Our starting point of the security estimation presented here is to assume that the quantum state effectively shared by Alice, Bob and Eve, is of product form $\rho_{ABE}^{\otimes n}$. In contrast to that, the most general coherent attacks can introduce correlations between the quantum states describing subsequent signals. However, it is known that this kind of attack does not give any advantage to Eve in the asymptotic key limit, if the local dimension of the involved Hilbert spaces are finite \cite{renner07a}. Unfortunately, the quantum de Finetti theorem cannot be directly applied here, as one needs to bound the local dimension of Bob's received states, which generally is infinite dimensional in CV-QKD. Recent work \cite{Christandl07suba} indicates that there is hope that one can extend this results to the infinite dimensional case.

The experimental feasibility of various CV-QKD schemes using coherent states as input and variations of homodyne detection has already been demonstrated \cite{grosshans03a,hirano03a,lorenz04a,lance05a,lorenz06a,symul07a,lodewyck07b}. Although promising from a technological point of view as the measurement can operate at high repetition rates, the efficiency of these schemes seems to be limited by the classical post-processing protocol. In general, the performance  can be improved by using reverse reconciliation (RR): one reverses the flow of classical information in the error-correction step of the protocol, so that the raw key is based upon Bob's measurement results \cite{grosshans03a}. If a practical error-correction procedure with non-ideal efficiency is considered, additional procedures like postselection  \cite{silberhorn02b} might become favorable to increase the efficiency \cite{heid06a}. Here, we limit ourselves to the idealized scenario of CV-QKD involving noiseless detectors and perfect error-correction. Consequently, we suppose that a RR protocol without postselection procedures is used. The aim is to present the still missing security analysis for a discrete modulated CV-QKD valid in an idealized setting but considering arbitrary noise in the collective attack scenario. It should also be noted that in a typical physical realization, Alice sends an additional phase reference pulse to Bob via Eve's domain. In general, Eve could interact with this additional mode as well to gain more information about the exchanged signals. As shown by H\"aseler \emph{et al.} \cite{haseler08a}, a full security proof would have to take the full two mode structure of the signals into account, but also additional measurements have to be done to test the reference-signal structure of the two modes. Here, we present a simplified proof and assume that Bob's phase reference is prepared locally. Consequently, our signals are single modes. 

Typical experiments show that the dominant contribution to the excess noise in CV-QKD is due to the electronic noise of the detectors \cite{lorenz06a}. Therefore, we expect the channel excess noise relevant in CV-QKD to be relatively low and of the order of a few percent. Our analysis is based on work done by Rigas \cite{rigas06b}, who gave an estimation of the maximal eigenvalues and corresponding eigenstates of a quantum state based on homodyne detection.  In our protocol, Alice uses coherent states as signals.  If the quantum channel imposes loss onto the signals, but is noiseless otherwise, Bob's received states $\rho_{B}^{x}$, conditioned on Alice sending the bit-value $x$, are pure coherent states. In contrast to that, Bob will receive mixed conditional states if the quantum channel imposes additional noise upon the signals. Consequently, the maximal eigenvalue of the received states $\rho^{x}_{B}$ will deviate from unity as $1-\tilde{\varepsilon}_{x}$. In this article, we use $\tilde{\varepsilon}_{x}$ together with the overlap of the corresponding eigenstates $\tilde{\varepsilon}_{x}$ as a figure of merit to quantify the amount excess noise present in the quantum channel. These parameters will be connected to the observed measurement outcomes of Alice and Bob in Sec. V. For $\tilde{\varepsilon}_{x}=0$, we retrieve the known results for the lossy channel given in Ref. \cite{heid06a}. Therefore, we expect our approach to yield positive key rates as long as the noise of the quantum channel and consequently $\tilde{\varepsilon}_{x}$ is small enough.

This article is organized as follows: In the next section, we introduce a binary CV-QKD protocol where Bob is allowed to coarse grain his continuous measurement outcomes to discrete bit-values arbitrarily, which will be used as the raw key. Therefore, we modify the known security analysis for collective attacks to include this additional step in Sec. III. Then, we proceed by computing the secret key rate of a binary CV-QKD protocol with a fixed discretization of the continuous measurement outcomes. This will be done in two steps: in Sec. IV, we give an expression for the secret key rate in terms of maximal eigenvalues and corresponding eigenstates of Bob's received conditional states. These parameters are then estimated via Bob's homodyne measurement in the proceeding section. We conclude with a numerical evaluation of the secret key rate in a experimental relevant scenario and a discussion of the results. 

\section{The Protocol}

We consider a prepare-and-measure protocol using continuous variable states and homodyne detection. In general, we allow Bob to discretize his continuous measurement outcomes and to do announcements arbitrarily. However, we also give a description of a concrete protocol as an example with those steps specified. This specific protocol will be used in Sec. VI to evaluate the secret key rate for a typical experiment numerically.
Any QKD protocol can be decomposed into two phases. In the first phase Alice
prepares quantum states and sends them to Bob, who then performs measurements on
them. In the second phase, Alice and Bob use an authenticated two-way
channel for classical communication to turn the classical data (knowledge of
signals sent, and measurement results) into a secret key.

\begin{description}
\item[Quantum phase:]
$\ $\\
\begin{itemize}
\item[1.]  Alice sends a sequence of coherent states with amplitude $\alpha$ but
randomly selected opposing phase, $|\alpha \rangle $ or $|-\alpha \rangle $,
to Bob. Alice stores her choice for signal $i$ in a variable $x_i$ by
assigning to the choice $|\alpha \rangle$ the value $x_i=1$, and to $| -
\alpha \rangle$ to $x_i=0$.

\item[2.] Bob randomly measures each signal with a homodyne measurement
corresponding to the $q$ or $p$ quadratures \cite{silberhorn02b}. We denote
Bob's measurement results as $y_i$ and denote the basis choice by the binary
variable $b_i$ (We choose the reference frame such that the signal states
are modulated in the $q$ quadratures.
\end{itemize}
\item[Classical phase:]
$\ $\\
\begin{itemize}
\item[3.] After the quantum phase, Bob announces for each signal the measurement
basis.

\item[4.] Alice and Bob test their correlations by publishing randomly selected data points $x_{i}$ and $y_{i}$. Moreover all of their data  (Alice's modulation and Bob's full measurement result) that originated from Bob measuring the $p$ quadrature is published and used to check for Eve's interference.

\item[5.] Alice and Bob dismiss the data that originated from measuring in the $p$ basis for the remaining key distillation part of the protocol in order to obtain the sifted key.

\item[6.] Let us denote the string of outcomes pertaining to the sifted key as $\{\vec{x},\vec{y}\}$. From the collection of outcomes $\vec{y}$ Bob computes a string $\vec{u}$ and $\vec{\tilde{y}}$ to that we will refer to as the \emph{announcement} and the \emph{discretization} in the following.

\item[7.] Bob announces $\vec{u}$ and keeps $\vec{\tilde{y}}$.  In general, the announced vector $\vec{u}$ will have continuous entries. Bob could, for example, announce the modulus $|y_{i}|$ of his measurement result, whenever he chose the $q$-quadrature as basis. The discretization $\vec{\tilde{y}}$ is vector with discrete entries from which the secret key will be generated. This could be, for example, the sign of Bob's outcomes $y_{i}$ whenever he measured the $q$-quadrature.

\item[8.] Bob sends Alice error correction information to allow her to
reconcile her string $\vec{x}$ of the sifted data to the corresponding
string $\vec{\tilde{y}}$.

\item[9.] Alice and Bob do privacy amplification by applying universal-2 hash
functions to the  string $\vec{\tilde{y}}$, now shared by Alice and Bob. This will effectively shorten the string $\vec{\tilde{y}}$ by $n\tau$ bits of information, where $n$ is number of transmitted signals.
\end{itemize}
\end{description}

This protocol is equivalent to an entanglement based protocol \cite{bennett92c}. In step 1, Alice prepares a entangled state $|\Psi \rangle =\frac{1}{\sqrt{2}}(|0\rangle |-\alpha \rangle +|1\rangle|\alpha \rangle )$ and sends the coherent state system to Bob. Then she measures her state in the $|0\rangle$ and $|1\rangle$ basis. Steps 2 to 9 remain the same.

\section{The secret key rate in the infinite key limit}

 Our security analysis follows the one given in Ref. \cite{renner05b,devetak05a}. Here, we limit ourselves to the asymptotic key limit as the number of entries $n$ in the raw key $\vec{y}$ tend to infinity. Therefore, we only consider leading terms in $n$ in the formulas. Let $\mathbf{X}$, $\mathbf{Y}$, $\mathbf{\tilde{Y}}$ and $\mathbf{U}$ denote random variables that can take the values $\vec{x}$, $\vec{y}$, $\vec{\tilde{y}}$, $\vec{u}$ as introduced in the preceding section.
In step 6 of our protocol, Bob announces $\vec{u}$, so that this information becomes available to both Alice and Eve. The classical information contained in the announcement can be formally embedded in a quantum system
$\rho_{\mathbf{U}}$. After the announcement, the system $\rho_{\mathbf{XU}}$ describes Alice's data and $\rho _{E
\mathbf{U}}$ describes the state Eve holds. Later in the step 8 of
the protocol Bob sends error correction information over the public channel to Alice. As Eve can listen to this channel, the information $W$ about the key contained in the error correction becomes available to her. Again, we can formally embed this classical information in a quantum state $\rho_{\mathbf{W}}$. After the error correction, Alice and Bob share $\vec{\tilde{y}}$ and Eve's knowledge about the
exchanged data is summarized in a state $\rho _{E \mathbf{UW}}$.
According to Ref. \cite{renner05a} one has to shrink the raw key by
$n \tau=S(\mathbf{\tilde{Y}}:E\mathbf{UW})$ bits of information in
the asymptotic key limit, where $S$ denotes the quantum mutual
information \cite{nielsen00a}, so that the final key will be secure
with high probability. The secret key rate that Alice and Bob
finally can obtain is given by $H(\mathbf{\tilde{Y}})-n \tau$, where
$H(\mathbf{\tilde{Y}})$ describes the Shannon entropy of
$\mathbf{\tilde{Y}}$, which can be evaluated after the channel test.
From Ref. \cite{renner05b} we know that
\begin{equation}
n \tau=S(\mathbf{\tilde{Y}}:E\mathbf{UW})\leq
S(\mathbf{\tilde{Y}}:E\mathbf{U})+I(\mathbf{\tilde{Y}}:\mathbf{W}),
\label{Icut}
\end{equation}
where $I$ denotes the Shannon mutual information \cite{shannon48a}.
Alice has to correct all the errors in her string $\vec{x}$ in step
8 of the protocol. Therefore, Bob sends Alice error correction
information. The amount of error correction information necessary
for Alice to succeed is given by
\begin{equation}
I(\mathbf{\tilde{Y}}:\mathbf{W})=f(e)[H(\mathbf{\tilde{Y}})-I(\mathbf{XU}:\mathbf{\tilde{Y}})],  \label{Ierror}
\end{equation}%
where $f(e)\geq 1$ denotes the efficiency of the error correction procedure.
Alice and Bob know the amount of published error correction information
after step 8. In the following, we assume that the error correction is
ideal, so that $f(e)=1$. From the Eqs. (\ref{Icut},\ref{Ierror}) we know
that we have to shrink the key in the privacy amplification step by
\begin{equation*}
n \tau\leq S(\mathbf{\tilde{Y}}:E\mathbf{U})+H(\mathbf{\tilde{Y}})-I(\mathbf{XU}:\mathbf{\tilde{Y}}),
\end{equation*}
bits of information. The length of the final secret key that Alice
and Bob can obtain is given by
\begin{eqnarray}  \label{e1}
n G &=&H(\mathbf{\tilde{Y}})-n \tau   \\
&\geq&I(\mathbf{XU}:\mathbf{\tilde{Y}})-S(E\mathbf{U}:\mathbf{\tilde{Y}})  \nonumber\\
&=&I(\mathbf{X}:\mathbf{\tilde{Y}}|\mathbf{U})-S(E:\mathbf{\tilde{Y}}|\mathbf{U})\nonumber\;.
\end{eqnarray}
In the third line we have used the result that
$S(UV:W)=S(U:W|V)+S(V:W)$, which also holds for the classical mutual
information $I(UV:W)$ in particular. The length of the secret key
can be lower bounded as
\begin{eqnarray}
n G &=&I(\mathbf{X}:\mathbf{\tilde{Y}}|\mathbf{U})-S(E:\mathbf{\tilde{Y}}|\mathbf{U})  \notag \\
&=& I(\mathbf{X}:\mathbf{\tilde{Y}}|\mathbf{U})-S(E|\mathbf{U})+S(E|\mathbf{U\tilde{Y}}) \notag\\
&\geq&I(\mathbf{X}:\mathbf{\tilde{Y}}|\mathbf{U})-S(\mathbf{Y}:E)\;,  \label{e2a}
\end{eqnarray}%
where we have used the definition of the quantum mutual information $S(E:\mathbf{\tilde{Y}}|\mathbf{U})$ in the second line. The third line follows from the concavity of the entropy \cite{nielsen00a} as we will explain now. After Alice's and Bob's measurements, Eve's knowledge about the exchanged data is summarized in conditional quantum states $\rho_{E}^{\vec{x},\vec{y}}$.
Eve's states conditioned on Bob's measurement outcomes $y$ are therefore given by
\begin{equation}
  \rho_{E}^{\vec{y}}=\sum_{\vec{x}}P(\vec{x}|\vec{y})\rho_{E}^{\vec{x},\vec{y}}\;.
\end{equation}
From the measured outcomes $\vec{y}$, Bob computes the announcement $\vec{u}$ and the discretization $\vec{\tilde{y}}$. This can be modelled by a classical channel described by some given conditional probability distribution $P\left(\vec{y}|\vec{u}\right), \vec{\tilde{y}}$. The state $\rho_{E}^{\vec{u},\vec{\tilde{y}}}$ can therefore be written as
\begin{equation}\label{rhoEuytilde}
  \rho_{E}^{\vec{u},\vec{\tilde{y}}}=\sum_{\vec{y}}P\left(\vec{y}|\vec{u},\vec{\tilde{y}}\right) \rho_{E}^{\vec{y}}\;.
\end{equation}
It follows that the conditional entropy $S(E|\mathbf{U\tilde{Y}})$ can be bounded from below as
\begin{align}\label{schnick}
  S(E|\mathbf{U\tilde{Y}})&=\sum_{\vec{\tilde{y}}}\int d\vec{u} P(\vec{u},\vec{\tilde{y}}) S\left(\rho_{E}^{\vec{u},\vec{\tilde{y}}}\right)\\
&= \sum_{\vec{\tilde{y}}}\int d\vec{u} P(\vec{u},\vec{\tilde{y}}) S\left(\int d\vec{y} P\left(\vec{y}|\vec{u},\vec{\tilde{y}}\right) \rho_{E}^{\vec{y}}\right)\nonumber\\
&\geq \sum_{\vec{\tilde{y}}}\int d\vec{u} P(\vec{u},\vec{\tilde{y}}) \int d\vec{y} P\left(\vec{y}|\vec{u},\vec{\tilde{y}}\right)S(\rho_{E}^{\vec{y}})\nonumber\\
&= \sum_{\vec{y}} P(\vec{y}) S(\rho_{E}^{\vec{y}})=S(E|\mathbf{Y})\nonumber\;,
\end{align}
where we first used Eq. (\ref{rhoEuytilde}) and then the concavity of the entropy.
Since the conditional entropy $S(E|\mathbf{U})$ obeys  $S(E|\mathbf{U})\leq S(E)$
by the concavity of the entropy \cite{nielsen00a}, the last line of Eq.(\ref{e2a}) follows with the help of Eq. (\ref{schnick}).

The lower bound in Eq. (\ref{e2a}) has two terms, one depending on
the discretization $\mathbf{\tilde{Y}}$, one independent of it. We expect
to be able to find a discretization for arbitrary correlations
between Alice and Bob, so that the first term goes to
$I(\mathbf{X}:\mathbf{Y})$, e.g. a family of discretizations
$\mathbf{\tilde{Y}}_{\Delta}$ that tend to the identity
$\mathbf{\tilde{Y}}_{\Delta}\rightarrow\mathbf{Y}$ asymptotically as
$\Delta\rightarrow 0$. Here, the parameter $\Delta$ describes the
size of the coarse-graining of continuous measurement outcomes to a
certain discrete value. In Sec. VI we will give a simple example of
a discretization that can achieve the bound
$I(\mathbf{X}:\mathbf{Y})$ for particular class of correlations
between Alice and Bob without an asymptotic procedure.

In the following, we limit our security analysis to the collective
attack scenario and assume that the total state shared by Alice, Bob and Eve has
tensor product form $\rho_{ABE}^{\otimes n}$. Thus, the measurement
outcomes $x_{i}$ and $y_{i}$ are independently identical
distributed, and we can limit ourselves to single letter
distributions. Then, Bob computes $\tilde{y}$ and announces values
$u$ from his measured value of $y$. Therefore, Eq. (\ref{e2a}) can
be simplified as
\begin{equation}
G \geq I(X:\tilde{Y}|U)-S(Y:E)\;,  \label{e2}
\end{equation}
where we have introduced the single letter random variables $X$, $Y$, $\tilde{Y}$ and $U$ that can take the values $x$, $y$, $\tilde{y}$ and $u$ respectively.
The remaining central problem is to find a upper bound to $S(E:Y)$ as the first term is already available from the observed outcomes.
Without loss of the generality, we can assume Eve holds the purification of $\rho _{ABE}$. Define the set $\Xi _{ABE}(\rho )$ as a collection
of all of the possible pure state $\rho _{ABE}$ that compatible with the
observations available from the measurement. The secret key rate is then given by
\begin{equation}\label{keyrate}
G\geq I(X:\tilde{Y}|U)-\underset{\rho _{ABE}\in \Xi _{ABE}(\rho )}{\max }S(Y:E).
\end{equation}

In this article, we calculate this expression (\ref{keyrate}) for the binary modulated CV-QKD scheme introduced in Sec. II. This will be done as follows: first, we will divide the entropy $S(Y:E)$ into three terms. Then we will give an upper bound to each term independently. These bounds can either be directly given by Bob's observed first and second moments or can be expressed as functions of
the maximal eigenvalues and corresponding eigenstates of Eve's conditional
states. We conclude our proof by estimating these parameters via the first
and second moments of Bob's homodyne measurements using the results of Ref.
\cite{rigas06b} combined with an argument based on Schmidt's decomposition. In
the last section we evaluate the expected secret key rate $G$ for typical observations numerically.

\section{Lower bound on the secret key rate}

The central problem of calculating the secret key rate in a reverse reconciliation scheme according to Eq. \ref{keyrate} is to find an upper bound for the mutual information $S(Y:E)$ that can be estimated by observable quantities. This will be done in the following. As the mutual information between Alice and Eve is given by
\begin{equation}
S(X:E)=S(E)-S(E|X) \;,  \notag
\end{equation}
one can express the quantum mutual information $S(Y:E)$ between Bob and Eve in Eq.(\ref{keyrate}) as
\begin{equation}
S(Y:E)=S(E|X)+S(X:E)-S(E|Y) \;.  \label{SYE}
\end{equation}
As already mentioned, we will proceed to calculate an upper bound for $S(Y:E)$ by bounding the three terms $S(E|X)$, $S(X:E)$ and $S(E|Y)$ on the right hand side of Eq. (\ref{SYE}) individually. As we will see later, we can directly compute an upper bound for $S(E|X)$ from Bob's observed data. The remaining two terms will be given as functions of the maximal eigenvalues $1-\tilde{\varepsilon}_{x}$ and corresponding eigenstates $|\tilde{\varepsilon}_{x}\rangle $ of Eve's conditional states $\rho _{E}^{x}$.

In Ref. \cite{rigas06b}, Rigas presented an estimation of the
maximal eigenvalue and corresponding eigenstate of an unknown quantum state
based on the first and second moments of a homodyne measurement. We use this
result to estimate the biggest eigenvalue $1-\tilde{\varepsilon}_{x}$ and corresponding eigenstate  $|\tilde{\varepsilon}_{x}\rangle $ of Eve's conditional states  $\rho _{E}^{x}$ via Bob's measurements. We can express Eve's conditional states using this notation as
\begin{equation}  \label{decompE}
{\rho }_{E}^{x}=(1-\tilde{\varepsilon}_{x})|\tilde{\varepsilon}_{x}\rangle     \langle
\tilde{\varepsilon}_{x}|+\tilde{\varepsilon}_{x}\sigma _{E}^{x} \;,
\end{equation}
where $|\tilde{\varepsilon}_{x}\rangle \langle \tilde{\varepsilon}_{x}|$ have $\sigma _{E}^{x}$
orthogonal support. We will refer to the eigenstate belonging to the maximal eigenvalue as the maximal eigenstate.

In the following, we will assume that the maximal eigenvalues $1-\tilde{\varepsilon}_{x}$ and eigenvectors $|\tilde{\varepsilon}_{x}\rangle$ are given. Section V contains an estimation of these parameters from measurement data and will conclude our approach.

It turns out that an upper bound for Eve's conditional entropy $S(E|X)$, the first term on the right hand side
of Eq. (\ref{SYE}), can be obtained by exploiting Gaussian extremality
properties \cite{wolf06a}. The second term is the mutual information between Alice and Eve $S(X:E)$, which can be upper bounded by employing a
suitable purification method. The estimation of the third term, the entropy $S(E|Y)$ conditioned on Bob's measurement outcomes $Y$ is technically more involved and includes a linearization of the respective quantities, so that a bound can be evaluated.

\subsection{Eve's entropy $S(E|X)$  conditioned on Alice's variable $X$}

For given first and second moments of Bob's measurement outcomes, we have to find an upper bound for Eve's conditional entropy $S(E|X)$, which is the first term on the right hand side of Eq. (\ref{SYE}). The \emph{a priori} probabilities
$P(x)$ are fixed by Alice's state preparation. In the entanglement based
description of the protocol, Alice's state preparation is equivalent to
projection measurement onto her $A$ system of a pure three party state $\rho_{ABE}$. It follows that the combined two party state $\rho_{EB}^x=|\Psi
_{BE}^{x}\rangle\langle\Psi _{BE}^{x}|$ between Eve and Bob conditioned on
Alice's measurement outcome $x$ is pure. Therefore, by Schmidt's
decomposition, we conclude that $S({\rho }_{E}^{x})=S({\rho }_{B}^{x})$ \cite{nielsen00a}. It is known that the state with maximal entropy $S({\rho }_{B}^{x})$ for fixed first and second moments is Gaussian \cite{wolf06a,agarwal71a}. Since $S({\rho }_{E}^{x})=S({\rho }_{B}^{x})$ and $P(x)$ is fixed, one can directly apply the result given in Eqs. (15) and (16) of Ref. \cite{agarwal71a},
so that
\begin{align}  \label{SE|X}
S(E|X)&=\sum_{x}P(x)S(\rho _{E}^{x})   \\
&\leq \frac{1}{2}\sum_{x}[(1+V_{x})\log _{2}(1+V_{x})-V_{x}\log _{2}V_{x}]
\;.\nonumber
\end{align}
The term
\begin{equation}\label{defVx}
V_{x}=\sqrt{V_{Y_{q|x}}^{2}V_{Y_{p|x}}^{2}}-1/2
\end{equation}
quantifies the amount of excess noise imposed by the quantum channel connecting Alice and Bob. It is a function of Bob's observed
variances $V_{Y_{q}|X}^{2}$ and $V_{Y_{p}|X}^{2}$ of the corresponding quadrature distributions, that are given by
\begin{align}\label{quadvar}
  V_{Y_{q}|X}^{2}&=\mathrm{tr} \left(\rho_{B}^{x} \hat{q}^{2}\right)- \left[\mathrm{tr}\left(\rho_{B}^{x} \hat{q} \right)\right]^{2}\\
  V_{Y_{p}|X}^{2}&=\mathrm{tr} \left(\rho_{B}^{x} \hat{p}^{2}\right)- \left[\mathrm{tr}\left(\rho_{B}^{x} \hat{p} \right)\right]^{2}\;,\nonumber\\
\end{align}
and the quadrature operators $\hat{q}$ and $\hat{p}$ are defined as
\begin{align}\label{quad}
  \hat{q}&=\frac{1}{\sqrt2}\left(\hat{a}+\hat{a}^{\dagger}\right)\\
  \hat{p}&=\frac{\mathrm{i}}{\sqrt2}\left(\hat{a}-\hat{a}^{\dagger}\right)\;,\nonumber
\end{align}
whereas $\hat{a}$ and $\hat{a}^{\dagger}$ denote the photon annihilation and creation operators.

\subsection{The mutual information $S(X:E)$ between Alice and Eve}

Here, we employ methods known from state estimation to calculate the mutual information term $S(X:E)$ between Alice and Eve in Eq. (\ref{SYE}). After interacting with the signal states, Eve holds the conditional states $\rho_{E}^{x}$ in her ancilla system, that she wants to distinguish optimally in order to maximize the mutual information $S(X:E)$. If we introduce an auxiliary system $Q$ that contains a purification of the states $\rho_{E}^{0}$ and $\rho_{E}^{1}$, we can give an upper bound for $S(X:E)$: the mutual information can never increase when discarding subsystems, so that
\begin{equation}\label{incmut}
S(X:E)\leq S(X:QE)
\end{equation} 
holds. We choose the purification $Q$, so that the conditional states $\rho_{E}^{x}$ are purified as $|\Psi_{EQ}^{x}\rangle$. There are certainly purifications that would leak too much information to Eve, i.e. if one would supply Eve with a purification of the global state $\rho_{XQE}$. Since Eq.(\ref{incmut}) is valid for any purification, we would ideally choose one that minimizes $S(X:QE)$ to make the bound (\ref{incmut}) as tight as possible. This problem is closely connected to Uhlmann's theorem, as we will show now.

It has been shown that the quantum mutual information between a classical register described by the binary variable $X$ and a quantum system $QE$ can be expressed as
\begin{equation}\label{binS}
S(X:QE)=h\left[\frac{1}{2}\left(1-| \langle \Psi_{EQ}^{0}\left|\Psi_{EQ}^{1}\rangle \right|\right)\right],
\end{equation}
if the conditional states $|\Psi_{EQ}^{x}\rangle$ are pure \cite{heid06a}.
Here, $h$ denotes the binary entropy function
\begin{equation}
  h(z)=-z \log_{2} z -(1-z) \log_{2}(1-z)\;.
\end{equation}
Since $S(X:EQ)$\ monotonously increases with decreasing overlap $|\langle \Psi _{EQ}^{0}|\Psi _{EQ}^{1}\rangle |$, it is sufficient to find the purification $Q$ that maximizes the overlap $|\langle \Psi _{EQ}^{0}\left|\Psi _{EQ}^{1}\rangle \right|$ to minimize $S(X:EQ)$. The solution to this problem is known as Uhlmann's theorem \cite{nielsen00a}: 
\begin{equation}\label{uhltheo}
F\left(\rho_{E}^{0},\rho_{E}^{1}\right)=\max_{|\Psi_{EQ}^{0}\rangle,|\Psi_{EQ}^{1}\rangle}|\langle \Psi _{EQ}^{0}|\Psi _{EQ}^{1}\rangle |
\end{equation}
Here, the Uhlmann fidelity $F\left(\rho_{E}^{0},\rho_{E}^{1}\right)$ is defined as
\begin{equation}\label{uhlfid}
F\left(\rho_{E}^{0},\rho_{E}^{1}\right)=\mathrm{tr_{E}}\left(\sqrt{\sqrt{\rho_{E}^{0}}\rho_{E}^{1}\sqrt{\rho_{E}^{0}}}\right)\;.
\end{equation}
Therefore, we conclude that the tightest bound obtainable from Eq. (\ref{incmut}) to mutual information $S(X:QE)$ for a binary modulated setup is given by Eq. (\ref{uhltheo}) and Eq. (\ref{binS}) as
\begin{equation}\label{upboundXE}
  S(X:E)\leq h\left[\frac{1}{2}\left\{1-F\left(\rho_{E}^{0},\rho_{E}^{1}\right)\right\}\right]
\end{equation} 

In general, the upper bound (\ref{upboundXE}) of the mutual information $S(X:E)$ can be calculated, if the Eve's conditional states $\rho_{E}^{x}$ are known. However, the full information about the states $\rho_{E}^{x}$ is usually not available from measurements. As already mentioned, we base our security analysis on the estimation of the maximal eigenvalues $1-\tilde{\varepsilon}_{x}$ and corresponding eigenstates $|\tilde{\varepsilon}_{x}\rangle$ of Eve's conditional states $\rho_{E}^{x}$ that we will estimate by Alice and Bob's observation. Therefore, we proceed by giving an upper bound of $S(X:E)$ as function of these parameters. This can be done by by considering a particular purification $Q$.

Any purification $|\Psi _{EQ}^{x}\rangle $ can be expanded as
\begin{equation}  \label{decompQ}
|\Psi _{EQ}^{x}\rangle =\sum_{i}c_{i}^{x}|i_{Q }^{x}\rangle|i_{E}^{x}\rangle \;.
\end{equation}
Without loss of generality, we can choose the first term in the Schmidt-decomposition (\ref{decompQ}) to correspond to the maximal eigenvalue ${c_{0}^{x}}^2:=1-\tilde{\varepsilon}_{x}$. The corresponding eigenstate is then given by Eq. (\ref{decompE}) as $|0_{E}^{x}\rangle=|\tilde{\varepsilon}_{x}\rangle$. With the help of expansion (\ref{decompQ}), the modulus of the overlap between the two conditional states can be evaluated as
\begin{equation}
\left|\langle \Psi _{EQ}^{0}|\Psi _{EQ}^{1}\rangle \right|=\left|\sum_{ij}c_{i}^{0}c_{j}^{1}\langle i _{i}^{0}|j_{Q}^{1}\rangle \langle i
_{E}^{0}|j _{E}^{1}\rangle \right| \;.
\end{equation}
If one chooses $\langle i _{Q}^{0}|j_{Q}^{1}\rangle =\delta _{ij}e^{i\varphi _{i}}$, where $\delta _{ij}$
is the Kronecker delta function and the phase $\varphi _{i}$ is the negative
of the phase of the complex number $\langle i_{E}^{0}|i _{E}^{1}\rangle $, it follows that
\begin{eqnarray}\label{ovE}
\left|\langle \Psi _{EQ}^{0}|\Psi _{EQ}^{1}\rangle \right| &=&\left|\sum_{i}c_{i}^{0}c_{i}^{1}e^{i\varphi _{i}}\langle i_{E}^{0}|i _{E}^{1}\rangle
\right| \\
&\geq &\sqrt{(1-\tilde{\varepsilon}_{0})(1-\tilde{\varepsilon}_{1})}\left|\langle
\tilde{\varepsilon}_{0}|\tilde{\varepsilon}_{1}\rangle \right|\nonumber.
\end{eqnarray}

Therefore, we obtain a lower bound on the quantum mutual information $S(X:E)$ using Eq. (\ref{binS}) and Eq. (\ref{ovE}) as
\begin{align}  \label{SXE}
S(X:E)&\leq S(X:QE)\\&\leq h\left[\frac{1}{2}(1-\sqrt{(1-\tilde{\varepsilon}_{0})(1-\tilde{\varepsilon}_{1})}\gamma )\right]\;,\nonumber
\end{align}
where we introduced 
\begin{equation}  \label{defgamma}
\gamma :=\left|\langle\tilde{\varepsilon}_{0}|\tilde{\varepsilon}_{1}\rangle \right|\;,
\end{equation}
as a short hand notation for the overlap of Eve's maximal eigenstates. In Sec. V we will estimate the values for $\tilde{\varepsilon}_{x}$ and $\gamma$ via Bob's homodyne measurements.

\subsection{Eve's entropy $S(E|Y)$ conditioned on Bob's measurement outcome $Y$}

\begin{figure}[tbp]
  \includegraphics[height=6cm,width=9cm]{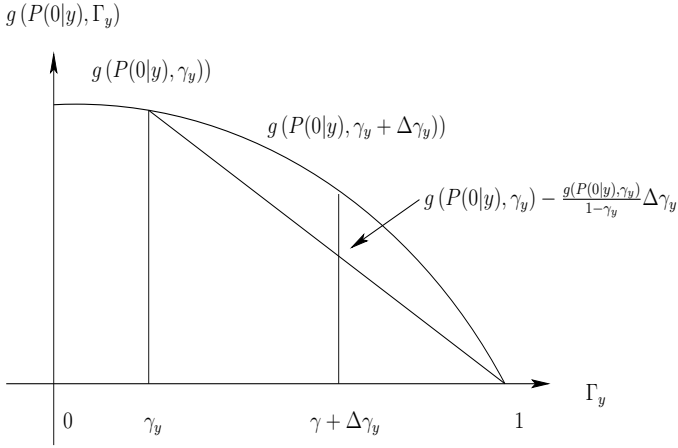}
  \caption{Schematical representation of the function
    $g\left(P(0|1),\Gamma_{y}\right)$.The validity of Eq.
    (\ref{approxg}) can easily be checked for all $\Gamma_{y}\leq
    \gamma_{y}+\Delta\gamma_y$.} 
  \label{gfuncpic}
\end{figure}

The last term of Eq. (\ref{SYE}) to be estimated reads
\begin{equation}  \label{SE|Y}
S(E|Y)=\int dyP(y)S(\rho _{E}^{y})\;.
\end{equation}
Prior to Alice's measurement, the three party state $\rho_{ABE}$ can be
assumed to be pure. Since Alice performs a projection measurement on
her subsystem, it follows that the combined two party state $%
\rho_{EB}^x=|\Psi _{BE}^{x}\rangle\langle\Psi _{BE}^{x}|$ between Eve and
Bob conditioned on Alice's measurement result is pure. 
Moreover, Bob performs a projection measurement $|y\rangle \langle y|$ on his
subsystem, so that Eve's state $|\Psi _{E}^{xy}\rangle$ conditioned on Alice's measurement outcome $x$ and Bob's outcome $y$ is pure.
Eve's states $\rho _{E}^{y}$ conditioned on Bob's measurement outcome $y$
can be written as
\begin{equation}
\rho _{E}^{y}=P(0|y)|\Psi _{E}^{0y}\rangle \langle \Psi
_{E}^{0y}|+P(1|y)|\Psi _{E}^{1y}\rangle \langle \Psi _{E}^{1y}|\;.
\end{equation}
From Sec. III.C of Ref. \cite{heid06a} we know that
\begin{align}
S(\rho _{E}^{y})&=h\left[\frac{1}{2}-\frac{1}{2}\sqrt{1-4P(0|y)P(1|y)(1-|\langle
\Psi _{E}^{0y}|\Psi _{E}^{1y}\rangle |^{2})}\right]  \notag \\
&=g\left(P(0|y),\left|\langle \Psi _{E}^{0y}|\Psi _{E}^{1y}\rangle \right|\right),  \label{SE|y}
\end{align}
where we have introduced the function $g\left(P(0|y),\left|\langle \Psi _{E}^{0y}|\Psi_{E}^{1y}\rangle\right|\right)$ as a shorthand notation. As we can see, the entropy $S(E|Y)$ to be evaluated is a function of the overlaps
\begin{equation}\label{DefGamma}
  \Gamma_{y}=\left|\langle \Psi _{E}^{0y}|\Psi _{E}^{1y}\rangle\right|\;,
\end{equation}
that depend on the outcomes $y$. Additionally, the probability distributions $P(0|y)$ and $P(y)$ need to be estimated by the channel test. We will proceed to lower bound the entropy $S(E|Y)$ (\ref{SE|Y}) by exploiting special properties of the $g$ function given by equation (\ref{SE|y}). It can be easily verified that $g\left(P(0|y),\Gamma_{y}\right)$ as a function of the overlaps has the following properties:
\begin{align}
  g(P(0|y),1)&=0\label{propg}\\
  \frac{\partial g(P(0|y),x)}{\partial x}&\leq 0\label{monoton}\\
  \frac{\partial^{2} g(P(0|y),x)}{\partial x^{2}}&\leq 0\label{concave}
\end{align}
We introduce positive and real parameters $\gamma_{y}$ and $\Delta\gamma_y$ such that we can rewrite the overlap $\Gamma_y$ (\ref{DefGamma}) as
\begin{equation}\label{decompGamma}
  \Gamma_{y}\leq \gamma_{y}+\Delta\gamma_y\;.
\end{equation}
It follows that for any $0\leq \Gamma_{y}\leq 1$ the inequality
\begin{align}\label{approxg}
  g(P(0|y),\Gamma_{y})&\geq g(P(0|y),\gamma_{y}+\Delta \gamma_{y})\\
  &\geq g(P(0|y),\gamma_{y})-\frac{g(P(0|y),\gamma_{y})}{1-\gamma_{y}}\Delta\gamma_{y}\nonumber
\end{align}
holds, as the first line of Eq. (\ref{approxg}) follows from the monotonicity (\ref{monoton}) and the second line follows from the concavity (\ref{concave}) together with property (\ref{propg}) if $0\leq \gamma_{y}\leq 1$. Later we will give explicit expressions for the decomposition (\ref{decompGamma}), so that these properties can easily be checked. Fig. (\ref{gfuncpic}) illustrates Eq. (\ref{approxg}) schematically.

Moreover, the approximation of Eq. (\ref{approxg}) can simplified further, if one could find a parameter $\tilde{\gamma}$ independent of $y$ with the properties $\tilde{\gamma}\geq\gamma_{y}$ and $\tilde{\gamma}\leq 1$, as
\begin{align}\label{approxg2}
  g(P(0|y),\gamma_{y})&\geq g\left(P(0|y),\tilde{\gamma}\right)\\
  \frac{g\left(P(0|y),\gamma_{y}\right)}{1-\gamma_{y}}&\leq \frac{g\left(P(0|y),\tilde{\gamma}\right)}{1-\tilde{\gamma}}\nonumber\;.
\end{align}
 We will see later that setting $\tilde\gamma$ to $\gamma$ as defined in Eq. (\ref{defgamma}) satisfies these constraints. The first bound of (\ref{approxg2}) is a simple consequence of the monotonicity (\ref{monoton}), whereas the second inequality follows from the properties (\ref{propg}--\ref{concave}). It can easily be verified by realizing that the quantity  $\frac{g\left(P(0|y),\Gamma_{y}\right)}{1-\Gamma_{y}}$ is given by the modulus of the gradient of the straight line connecting the points $g\left(P(0|y),\Gamma_{y}\right)$ and $g\left(P(0|y),\Gamma_{y}=1\right)=0$. From Fig. \ref{gfuncpic} it is obvious that this modulus increases if one chooses the point $\Gamma_{y}$ to be closer to one. Therefore, the second bound of (\ref{approxg2}) is valid for all $\tilde{\gamma}$ satisfying $\gamma_{y} \leq \tilde{\gamma} \leq 1$. Finally, we can estimate the conditional entropy $S(E|Y)$ given by Eq. (\ref{SE|Y}) with the help of the expressions (\ref{approxg}) and (\ref{approxg2}) as
\begin{align}  \label{e6}
S(E|Y)=&\int dy P(y)S(\rho _{E}^{y}) \\
\geq& \int dy P(y)g(P(0|y),\tilde{\gamma})  \nonumber \\
&-\frac{1}{1-\tilde{\gamma} }\int dy P(y)g(P(0|y),\tilde{\gamma})\Delta\gamma_{y}  \nonumber\;.\\
=& \int dy P(y)g(P(0|y),\tilde{\gamma})-\Delta S\nonumber\;,
\end{align}
where we introduced the term $\Delta S$ as a shorthand notation.

In the following, we will give explicit expressions for the missing parameters $\gamma_{y}$, $\Delta\gamma_{y}$ and $\tilde{\gamma}$ in order to connect these parameters to quantities that are observable to Alice and Bob. The starting point of this analysis is again noticing that the state $|\Psi _{BE}^{x}\rangle$ that Bob and Eve share conditioned on Alice's measurement outcome $x$ is pure, so that one can decompose it as
\begin{equation}
|\Psi _{BE}^{x}\rangle =\sqrt{(1-\tilde{\varepsilon}_{x})}|\tilde{\beta}_{x}\rangle |\tilde{\varepsilon}_{x}\rangle +\sqrt{\tilde{\varepsilon}_{x}}|\varphi _{EB}^{x}\rangle \;,  \label{stateBE}
\end{equation}
using Schmidt's decomposition theorem \cite{nielsen00a}. We have introduced eigenstate $|\tilde{\beta}_{x}\rangle$ of Bob's conditional density matrix $\rho_{B}^{x}$ corresponding to the maximal eigenvalue $1-\tilde{\varepsilon}_{x}$. All terms orthogonal to $|\tilde{\beta}_{x}\rangle|\tilde{\varepsilon}_{x}\rangle$ are summed up in the term $|\varphi _{EB}^{x}\rangle$, such that $\langle\tilde{\beta}_{x}|\varphi^{x}_{EB}\rangle=0$ and $\langle\tilde{\varepsilon}_{x}|\varphi^{x}_{EB}\rangle=0$. From Eq. (\ref{stateBE}), one can construct Eve's states $|\Psi_{E}^{xy}\rangle$ conditioned on both Alice's and Bob's measurement outcomes as
\begin{equation}  \label{Exy}
|\Psi _{E}^{xy}\rangle =\frac{\sqrt{(1-\tilde{\varepsilon}_{x})}\langle
y|\tilde{\beta}_{x}\rangle |\tilde{\varepsilon}_{x}\rangle +\sqrt{\tilde{\varepsilon}_{x}%
}\langle y_{B}|\varphi_{EB}^{x}\rangle }{\sqrt{P(y|x)}}\;.
\end{equation}
by projecting Bob's system onto $|y\rangle_{B} \langle y|$. 
The conditional probabilities $P(y|x)$ are given by
\begin{equation}
P(y|x)=(1-\tilde{\varepsilon}_{x})|\langle y|\tilde{\beta}_{x}\rangle |^{2}+%
\tilde{\varepsilon}_{x}\left|\langle \varphi_{EB}^{x}|y\rangle_{B}\langle y|\varphi_{EB}^{x}\rangle\right|^{2}\;.
\end{equation}
By setting
\begin{equation}  \label{defayx}
a_{y}^{x}=\frac{\langle y|\tilde{\beta}_{x}\rangle }{\sqrt{P(y|x)}}
\end{equation}
and
\begin{equation}  \label{defbyx}
b_{y}^{x}=\frac{\sqrt{\langle \varphi_{EB}^{x}|y\rangle_{B} \langle y|\varphi_{EB}^{x}\rangle }}{\sqrt{P(y|x)}}\;,
\end{equation}
we can express Eq.(\ref{Exy}) as
\begin{equation}\label{psiexy}
|\Psi _{E}^{xy}\rangle =\sqrt{(1-\tilde{\varepsilon}_{x})}a_{y}^{x}|\tilde{\varepsilon}_{x}\rangle +\sqrt{\tilde{\varepsilon}_{x}}b_{y}^{x}|\varphi_{E}^{xy}\rangle ,
\end{equation}
where $|\tilde{\varepsilon}_{x}\rangle$ is orthogonal to $|\varphi_{E}^{xy}\rangle$. The normalized states $|\varphi^{xy}_{E}\rangle$ are given by Eqs. (\ref{Exy}), (\ref{defayx}), (\ref{defbyx}) and (\ref{psiexy}) as
\begin{equation}
|\varphi_{E}^{xy}\rangle=\left(\langle \varphi_{EB}^{x}|y\rangle_{B} \langle y|\varphi_{EB}^{x}\rangle\right)^{-\frac{1}{2}}\langle y_{B}|\varphi_{EB}^{x}\rangle\;. 
\end{equation}
Without loss of generality, we can choose $a_{y}^{x}$ and $b_{y}^{x}$ to be real. Moreover, from expansion (\ref{psiexy}) it is obvious that
\begin{equation}\label{approxa}
  \sqrt{1-\tilde{\varepsilon}_{x}}a_{y}^{x}\leq 1
\end{equation}
holds.  The overlap $\Gamma_{y}$ is given by Eq. (\ref{psiexy}) as
\begin{align} \label{overlap}
\Gamma_{y}=&\left|\langle \Psi _{E}^{0y}|\Psi _{E}^{1y}\rangle \right|\\
=&\left|\sqrt{(1-\tilde{\varepsilon}
_{0})(1-\tilde{\varepsilon}_{1})}a_{y}^{0}a_{y}^{1}\langle \tilde{\varepsilon}_{0}|\tilde{\varepsilon}_{1}\rangle \right.  \notag \\
&+\sqrt{(1-\tilde{\varepsilon}_{0})\tilde{\varepsilon}_{1}}
a_{y}^{0}b_{y}^{1}\langle \tilde{\varepsilon}_{0}|\varphi _{E}^{1y}\rangle \notag\\
&+\sqrt{(1-\tilde{\varepsilon}_{1})\tilde{\varepsilon}_{0}}b_{y}^{0}a_{y}^{1}\langle
\varphi _{E}^{0y}|\tilde{\varepsilon}_{1}\rangle  \notag \\
&\left.+\sqrt{\tilde{\varepsilon}_{0}\tilde{\varepsilon}_{1}}b_{y}^{0}b_{y}^{1}
\langle \varphi _{E}^{0y}|\varphi _{E}^{1y}\rangle \right|\notag \;,
\end{align}
so that Eq. (\ref{decompGamma}) follows from (\ref{overlap}) by triangle inequality with the parameters $\gamma_{y}$ and $\Delta \gamma_{y}$ defined as
\begin{align}\label{approxgammay}
  \gamma_{y}=&\left|\sqrt{(1-\tilde{\varepsilon}
_{0})(1-\tilde{\varepsilon}_{1})}a_{y}^{0}a_{y}^{1}\langle \tilde{\varepsilon}_{0}|\tilde{\varepsilon}_{1}\rangle\right|\\=&\left|\sqrt{(1-\tilde{\varepsilon}
_{0})(1-\tilde{\varepsilon}_{1})}a_{y}^{0}a_{y}^{1}\right|\gamma\nonumber\\
\Delta\gamma_{y}=&\left|\sqrt{(1-\tilde{\varepsilon}_{0})\tilde{\varepsilon}_{1}}
a_{y}^{0}b_{y}^{1}\langle \tilde{\varepsilon}_{0}|\varphi _{E}^{1y}\rangle \right.\nonumber\\
&+\sqrt{(1-\tilde{\varepsilon}_{1})\tilde{\varepsilon}_{0}}b_{y}^{0}a_{y}^{1}\langle
\varphi _{E}^{0y}|\tilde{\varepsilon}_{1}\rangle  \nonumber \\
&\left.+\sqrt{\tilde{\varepsilon}_{0}\tilde{\varepsilon}_{1}}b_{y}^{0}b_{y}^{1}
\langle \varphi _{E}^{0y}|\varphi _{E}^{1y}\rangle \right|\nonumber\;.
\end{align}
With the help of  Eq. (\ref{approxa}), the parameter $\gamma_{y}$ can be upper bounded as
\begin{equation}
   \gamma_{y}=\left|\sqrt{(1-\tilde{\varepsilon}_{0})(1-\tilde{\varepsilon}_{1})}a_{y}^{0}a_{y}^{1}\right|\gamma\leq \gamma \;,
\end{equation} 
so that we can set 
\begin{equation}\label{gammaeqtildegamma}
\tilde{\gamma}=\gamma
\end{equation} to satisfy $\gamma_{y}\leq \tilde{\gamma}$. Moreover, it can easily be checked that  $0\leq\gamma_{y}\leq\gamma\leq 1$ using property (\ref{approxa}).

In principle, we have now everything at hand to lower bound the conditional entropy $S(E|Y)$ according to Eq. (\ref{e6}). However, as we will see later, we can only estimate the overlap $\gamma=\left|\langle\tilde{\varepsilon}_{0}|\tilde{\varepsilon}_{1}\rangle\right|$ and eigenvalues $1-\tilde{\varepsilon}_{x}$ from Bob's measurements. As a consequence, the parameter $\Delta\gamma_{y}$ cannot be estimated by the observation and consequently the term $\Delta S$ in Eq. (\ref{e6}) cannot computed directly. Since $\Delta S$ is monotone in the parameter $\Delta\gamma_{y}$, it is again possible to lower bound the entropy $S(E|Y)$ by looking for a suitable upper bound for $\Delta\gamma_{y}$ which is a function of Bob's observable parameters. Here, we estimate the parameter $\Delta\gamma_{y}$ starting from the definitions (\ref{approxgammay}) as
\begin{align}\label{approxDeltagamma}
\Delta\gamma_{y}\leq&\left|\sqrt{(1-\tilde{\varepsilon}_{0})\tilde{\varepsilon}_{1}}a_{y}^{0}b_{y}^{1}\right|\left|\langle \tilde{\varepsilon}_{0}|\varphi _{E}^{1y}\rangle \right|\\
&+\left|\sqrt{(1-\tilde{\varepsilon}_{1})\tilde{\varepsilon}_{0}}a_{y}^{1}b_{y}^{0}\right|\left|\langle\tilde{\varepsilon}_{1}|\varphi _{E}^{0y}\rangle\right|  \nonumber \\
&+\left|\sqrt{\tilde{\varepsilon}_{0}\tilde{\varepsilon}_{1}}b_{y}^{0}b_{y}^{1}\right|\left|\langle \varphi _{E}^{0y}|\varphi _{E}^{1y}\rangle \right|\nonumber\\
\leq&\sqrt{\left(1-\tilde{\varepsilon}_{0}\right)\tilde{\varepsilon}_{1}}a_{y}^{0}b_{y}^{1}\sqrt{1-\left|\langle \tilde{\varepsilon}_{0}|\tilde{\varepsilon}_{1}\rangle \right|^{2}}\nonumber\\
&+\sqrt{\left(1-\tilde{\varepsilon}_{1}\right)\tilde{\varepsilon}_{0}}a_{y}^{1}b_{y}^{0}\sqrt{1-\left|\langle\tilde{\varepsilon}_{0}|\tilde{\varepsilon}_{1}\rangle\right|^{2}}\nonumber\\
&+\sqrt{\tilde{\varepsilon}_{0}\tilde{\varepsilon}_{1}}b_{y}^{0}b_{y}^{1}\nonumber\\
\leq&\sqrt{1-\gamma^{2}}\left(\sqrt{\tilde{\varepsilon}_{1}}b_{y}^{1}+\sqrt{\tilde{\varepsilon}_{0}}b_{y}^{0}\right)+\sqrt{\tilde{\varepsilon}_{0}\tilde{\varepsilon}_{1}}b_{y}^{0}b_{y}^{1}\nonumber
\end{align}
where we first used the triangle inequality. For the second inequality in Eq. (\ref{approxDeltagamma}) we used $\left|\langle \varphi _{E}^{0y}|\varphi _{E}^{1y}\rangle \right|\leq 1$ and
\begin{equation}\label{orthovec}
  \left|\langle{\Phi}|\tilde{\varepsilon}_{x}\rangle\right|^{2}+ \left|\langle{\Phi}|\varphi_{E}^{xy}\rangle\right|^{2}\leq 1\;,
\end{equation}
which is valid for any vector $|\Phi\rangle$ by orthogonality of the states $|\tilde{\varepsilon}_{x}\rangle$ and $|\varphi_{E}^{xy}\rangle$. In particular, we used
\begin{align}\label{orthoeps}
\left|\langle\tilde{\varepsilon}_{0}|\tilde{\varepsilon}_{1}\rangle\right|^{2}+\left|\langle\tilde{\varepsilon}_{0}|\varphi^{1y}_{E}\rangle\right|^{2}&\leq 1\\
\left|\langle\tilde{\varepsilon}_{1}|\tilde{\varepsilon}_{0}\rangle\right|^{2}+\left|\langle\tilde{\varepsilon}_{1}|\varphi^{0y}_{E}\rangle\right|^{2}&\leq 1\nonumber\;,
\end{align}
which follows from Eq. (\ref{orthovec}) by setting $|\Phi\rangle=|\varepsilon_{0}\rangle$ and $x=1$ for the first inequality or respectively $|\Phi\rangle=|\varepsilon_{1}\rangle$ and $x=0$ for the last inequality in Eq. (\ref{orthoeps}).
In the last step of Eq. (\ref{approxDeltagamma}), we used the definition (\ref{defgamma}) of $\gamma$ and the bound (\ref{approxa}).

With the expression (\ref{approxDeltagamma}), we can upper bound the term $\Delta S$ of Eq.(\ref{e6}) as
\begin{align}  \label{DeltaS}
\Delta S\leq&\sqrt{\frac{1+\gamma}{1-\gamma}}\int dy P(y)g(P(0|y),\gamma)\left(\sqrt{\tilde{\varepsilon}_{0}}b_{y}^{0}+\sqrt{\tilde{\varepsilon}_{1}}b_{y}^{1}\right) \\
&+ \frac{1}{1-\gamma}\int dy P(y)g(P(0|y),\gamma)\left(\sqrt{\tilde{\varepsilon}_{0}\tilde{\varepsilon}_{1}}b_{y}^{0}b_{y}^{1}\right).  \nonumber
\end{align}
These integrals can be estimated first applying the completeness relation  $\int dy|y\rangle \langle y|=I$ of Bob's homodyne measurement to the definition (\ref{defbyx}). It follows that
\begin{equation}
\int dy P(y|x)b_{y}^{x}{}^{2}=\int dy\langle \varphi _{EB}^{x}|y\rangle_{B}
\langle y|\varphi _{EB}^{x}\rangle=1 \;.   \label{e4}
\end{equation}
This condition on the parameters $b_{y}^{x}$ enables us to upper bound the remaining terms in Eq. (\ref{DeltaS}) as
\begin{eqnarray}  \label{blubber}
&&\int dy P(y)g(P(0|y),\gamma)\sqrt{\tilde{\varepsilon}_{x}}b_{y}^{x} \\
&\leq &\sqrt{\frac{\tilde{\varepsilon}_{x}}{2}\int dy P(y)\frac{g^{2}(P(0|y),\gamma)}{P(x|y)}},  \notag
\end{eqnarray}%
and
\begin{equation} \label{e10}
\int dy P(y)g(P(0|y),\gamma)\sqrt{\tilde{\varepsilon}_{0}\tilde{\varepsilon}%
_{1}}b_{y}^{0}b_{y}^{1}\leq \sqrt{\tilde{\varepsilon}_{0}\tilde{\varepsilon}%
_{1}} g\left(\frac{1}{2},\gamma\right)\;.
\end{equation}
with the help of the Cauchy-Schwarz-Buniakovsky inequality \cite{gradshteyn94a}. Details of this estimation can be found in Appendix \ref{extreme}.

Let us summarize our results. We can use Eq. (\ref{gammaeqtildegamma}) in Eq. (\ref{e6}) to bound the conditional entropy $S(E|Y)$ as
\begin{equation}  \label{ichweissnet}
S(E|Y)\geq  \int dy P(y)g\left(P(0|y),\gamma\right) - \Delta S \;.
\end{equation}
It follows from the inequalities (\ref{DeltaS}), (\ref{blubber}) and (\ref{e10}) that the term $\Delta S$ can be upper bounded as 
\begin{equation}
\Delta S \leq \tilde{\varepsilon}_{0}k_{0}+\tilde{\varepsilon}_{1}k_{1}+\sqrt{\tilde{\varepsilon}_{0}\tilde{\varepsilon}_{1}}\tilde{k}
\;,
\end{equation}
where we defined parameters $k_{x}$ and $\tilde{k}$ as
\begin{align} \label{e11}
{k_{x}}&=\sqrt{\frac{1+\gamma}{2(1-\gamma)}\int dy P(y)\frac{g^{2}\left(P(0|y),\gamma\right)}{P(x|y)}}  \\
{\tilde{k}}&=\frac{1}{1-\gamma }g\left(\frac{1}{2},\gamma \right)\nonumber \;.
\end{align}
Finally, a lower bound for the conditional entropy $S(E|Y)$ is therefore given by Eqs. (\ref{ichweissnet}) and (\ref{DeltaS}) as
\begin{align}\label{e15}
S(E|Y)\geq& \int dy P(y)g\left(P(0|y),\gamma\right)\\
&-\sqrt{\tilde{\varepsilon}_{0}}k_{0}-\sqrt{\tilde{\varepsilon}_{1}}k_{1}-\sqrt{\tilde{\varepsilon}_{0}\tilde{\varepsilon}_{1}}\tilde{k}  \;.\nonumber
\end{align}

\subsection{The mutual information $S(Y:E)$ between Bob and Eve}

We have shown that an upper bound for the mutual information $S(Y:E)$ between Bob and Eve is given by Eq. (\ref{SYE}), (\ref{SE|X}), (\ref{SXE}) and (\ref{e15}) as
\begin{align} \label{e12}
S(Y:E)=& S(E|X)+S(X:E)-S(E|Y) \\
\leq& \frac{1}{2}\sum_{x}[(1+V_{x})\log _{2}(1+V_{x})-V_{x}\log _{2}V_{x}] \nonumber \\
&+ h\left[\frac{1}{2}(1-\sqrt{(1-\tilde{\varepsilon}_{0})(1-\tilde{\varepsilon}_{1})}\gamma )\right] \nonumber\\
&- \int dy P(y)g[P(0|y),\gamma ]\nonumber\\
&+\sqrt{\tilde{\varepsilon}_{0}}k_{0}+\sqrt{\tilde{\varepsilon}_{1}}k_{1}+\sqrt{\tilde{\varepsilon}_{0}\tilde{\varepsilon}_{1}}\tilde{k} \nonumber\\
=& \frac{1}{2}\sum_{x}[(1+V_{x})\log _{2}(1+V_{x})-V_{x}\log _{2}V_{x}] \nonumber \\
&+ s(\tilde{\varepsilon}_{x},\gamma ) \;.
\end{align}
The first term in Eq. (\ref{e12}) can be directly computed from Bob's observed variances (\ref{quadvar}) using Eq. (\ref{defVx}). Here, we define the function $s(\tilde{\varepsilon}_{x},\gamma )$ to summarize all terms that depend on the maximal eigenvalues $1-\tilde{\varepsilon}_{x}$ and overlap $\gamma$ of the corresponding eigenstates of Eve's conditional states. The remaining problem is to estimate these parameters via Bob's homodyne measurement.

\section{Maximal eigenvalue and eigenstate}

We have already shown in the last section that the two party states $|\Psi
_{BE}^{x}\rangle$ conditioned on Alice's measurement outcome $x$ can be
chosen to be pure. Therefore, one can expand these conditional states using
the Schmidt-decomposition (\ref{stateBE}), so that the state $|\tilde
{\beta}_{x}\rangle|\tilde{\varepsilon}_{x}\rangle $ is orthogonal to $|\varphi_{EB}^{x}\rangle $. From Eq.(\ref{stateBE}) it follows that the $\rho_{E}^{x}$ and
$\rho_{B}^{x}$ have the same spectrum. Moreover, the eigenvectors of Bob's and
Eve's system are determined up to a global unitary operation on Eve's
system. According to Eq. (\ref{e12}), we need to estimate the modulus of the overlap of Eve's maximal eigenstates $|\tilde{\varepsilon}_{x}\rangle$ and the maximal eigenvalues $1-\tilde{\varepsilon}_{x}$. These parameters can be estimated from the first and second moments of Bob's measured data \cite{rigas06b}, as we will see in the following.

Suppose the fidelity between Bob's received conditional state $\rho_{B}^{x}$ and a pure coherent state $|\overline{\beta}\rangle $ satisfies
\begin{equation}  \label{fidelity}
\langle \overline{\beta}_{x}|\rho _{B}^{x}|\overline{\beta}_{x}\rangle=1-\varepsilon_{x}\;.
\end{equation}
The amplitude $\overline{\beta}_{x}$ is given by the first moments of Bob's homodyne measurement as
\begin{align}\label{defbetabar}
 \mathrm{Re}(\overline{\beta}_{x})&=\mathrm{tr}(\rho_{B}^{x} \hat{q})\\
 \mathrm{Im}(\overline{\beta}_{x})&=\mathrm{tr}(\rho_{B}^{x} \hat{p}) \nonumber \;.
\end{align}
The quadrature operators $\hat{q}$ and $\hat{p}$ are defined in Eq. (\ref{quad}). In the following, we will refer to the parameter $\varepsilon_{x}$ as the mixedness of Bob's conditional states.

It has been shown by Rigas \cite{rigas06b} that the mixedness $\varepsilon_{x}$ of the conditional states can be upper bounded from the outcomes of a homodyne measurement as
\begin{equation}\label{e16}
\varepsilon _{x}\leq
\frac{1}{2}\left[(V_{Y_{q}|x}^{2}+\frac{1}{2})(V_{Y_{p}|x}^{2}+\frac{1}{2})-1\right]=U_{x},
\end{equation}
where $V_{Y_{q}|x}$ and  $V_{Y_{p}|x}^{2}$ denote the variances of the $q$- and $p$-quadrature distributions (\ref{quadvar}) conditioned on Alice's variable $x$. The proof for the estimation (\ref{e16}) is given in Appendix (\ref{mix}).
Moreover, one can also estimate the overlap $|\langle \tilde{\beta}_{0}|\tilde{\beta}_{1}\rangle |$ of Bob's maximal conditional eigenstates as
\begin{equation}  \label{e17}
c_{l}(\tilde{\varepsilon}_{x},\varepsilon _{x},\kappa)\leq |\langle \tilde{\beta}_{0}|\tilde{\beta}_{1}\rangle |\leq c_{u}(\tilde{\varepsilon}%
_{x},\varepsilon _{x},\kappa) \;,
\end{equation}
if one assumes that the fidelity (\ref{fidelity}) is given. Here, the parameter $\kappa$ is given by the overlap of the coherent states corresponding to the mean values (\ref{defbetabar}) as
\begin{equation}\label{defkappa}
\kappa=\left|\langle \overline{\beta}_{0}|\overline{\beta}_{1}\rangle\right| \;.
\end{equation}
The detailed expression of $c_{l}(\tilde{\varepsilon}_{x},\varepsilon _{x},\kappa)$ and $c_{u}(\tilde{\varepsilon}_{x},\varepsilon _{x},\kappa)$\ can be seen in Appendix \ref{estovereigen}. 

The results (\ref{e16}) and (\ref{e17}) can be used to estimate the maximal eigenvalues and overlap $\gamma$ of the corresponding eigenstates of Eve's reduced density matrix. From the Schmidt decomposition (\ref{stateBE}) it follows that the eigenvalues of Bob's and Eve's reduced conditional density matrices are identical, so that
\begin{equation}
  \tilde{\varepsilon}_{x}\leq\varepsilon_{x}
\end{equation}
can easily be seen by expanding $\rho_{B}^{x}$ in its eigenbasis.  Moreover, Eve's attack should preserve the inner product \cite{heid06a}, so that $\langle -\alpha |\alpha\rangle =\langle \Psi _{BE}^{0}|\Psi _{BE}^{1}\rangle $. In Appendix \ref{estimation} we show that this allows us to bound the overlap $\gamma$ of Eve's eigenstates as
\begin{equation}
d_{l}\leq \gamma \leq d_{u} \;,  \label{e18}
\end{equation}%
where
\begin{equation}  \label{dl}
d_{l}=\frac{|\langle -\alpha |\alpha \rangle |-\sqrt{[\sqrt{(1-\tilde{\varepsilon}_{1})\tilde{\varepsilon}_{0}}+\sqrt{(1-\tilde{\varepsilon}_{0})\tilde{\varepsilon}_{1}}]^{2}+\tilde{\varepsilon}_{1}\tilde{\varepsilon}_{0}}}{\sqrt{(1-\tilde{\varepsilon}_{0})(1-\tilde{\varepsilon}_{1})}c_{u}(\tilde{\varepsilon}_{x},\varepsilon _{x},\kappa)}
\end{equation}
and
\begin{equation}  \label{du}
d_{u}=\frac{|\langle -\alpha |\alpha \rangle |+\sqrt{[\sqrt{(1-\tilde{\varepsilon}_{1})\tilde{\varepsilon}_{0}}+\sqrt{(1-\tilde{\varepsilon}_{0})\tilde{\varepsilon}_{1}}]^{2}+\tilde{\varepsilon}_{1}\tilde{\varepsilon}_{0}}}{\sqrt{(1-\tilde{\varepsilon}_{0})(1-\tilde{\varepsilon}_{1})}c_{l}(\tilde{\varepsilon}_{x},\varepsilon _{x},\kappa)}\;.
\end{equation}
The functions $c_{l}(\tilde{\varepsilon}_{x},\varepsilon _{x},\kappa)$ and $c_{u}(\tilde{\varepsilon}_{x},\varepsilon _{x},\kappa)$ are the extremal values of the overlap $\left|\langle\tilde{\beta}_0|\tilde{\beta}_{1}\rangle\right|$ of Bob's maximal eigenstates as defined in Eq. (\ref{e17}).

If the first and second moments of Bob's measurement outcomes are fixed, $U_{x}$ is given by Eq. (\ref{e16}). Therefore, the parameters $\tilde{\varepsilon}_{x}$ that are compatible with the observed data can vary between $0 \leq \tilde{\varepsilon}_{x}\leq \varepsilon_{x} \leq U_{x}$. In that respect, the quantities $\varepsilon_{x}$ and $\tilde{\varepsilon}_{x}$ are interior parameters that can only be bounded by the value of the observable quantity $U_{x}$. For any given value of $\varepsilon_{x}$, $\tilde{\varepsilon}_{x}$ and $\kappa$, the interval of compatible overlaps $|\langle \tilde{\beta}_{0}|\tilde{\beta}_{1}\rangle |$ according to Eq. (\ref{e17}) can be given. This in turn determines the interval of possible overlaps $\gamma$ via Eq. (\ref{e18}). The value for $\kappa$ is obtainable from the first moments of Bob's homodyne measurement, as can be seen from Eq. (\ref{defbetabar}). Finally, the secret key rate can be obtained by
\begin{align}
G \geq &I(X:\tilde{Y}|U)-\max_{\begin{array}{c}{0\leq\tilde{\varepsilon}_{x}\leq \varepsilon_{x}\leq U_{x}}\\{d_{l}\leq \gamma \leq d_{u}}
\end{array}}
\{s(\tilde{\varepsilon}_{x},\gamma )\label{final key rate}\\
 &-\frac{1}{2}\sum_{x}[(1+V_{x})\log _{2}(1+V_{x})-V_{x}\log _{2}V_{x}]\}\; \nonumber .
\end{align}
The maximum is taken over the interior parameters $\tilde{\varepsilon}_{x}$, $\varepsilon_{x}$ and $\gamma$ satisfying the bounds shown. These interior parameters can vary in intervals that are fixed by the values of $U_{x}$ and $\kappa$ that can be determined from the observation. As the $s$-function (\ref{e12}) contains details about Bob's measured data via the probability distributions $P(y)$ and $P(0|y)$, this additional information must be estimated from the measured data to analyze the secret key rate numerically for a given observation.

\section{Numerical results}

\begin{figure}[tbp]
  \includegraphics[height=6cm,width=9cm]{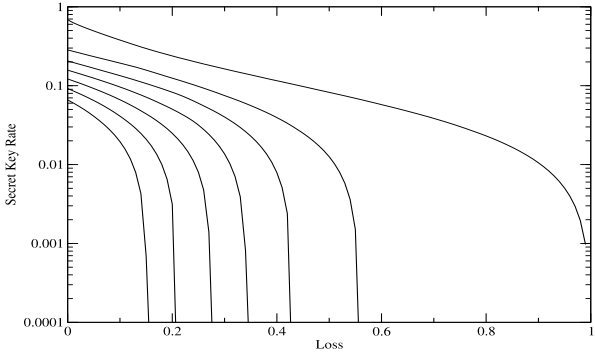}
  \caption{Secret key rate versus channel loss for a typical scenario with optimized signal strength. The different lines correspond to an excess noise $\delta$ of $\{0, 0.0004, 0.0008, 0.0012, 0.0016, 0.0020, 0.0024\}$.}  
  \label{figkey}
\end{figure}

The secret key rate (\ref{final key rate}) depends on Bob's observed probability distributions $P(y|x)$ directly via the mutual information term $I(X:\tilde{Y}|U)$ between Alice and Bob and via the term $s(\tilde{\varepsilon}_{x},\gamma)$, as can be seen from Eq. (\ref{e12}). The distribution  $P(y|x)$ is in principle available from experiments. To evaluate the secret key rate in an example, we simulate data for a typical experimental situation in which we find a Gaussian distribution \cite{grosshans03a,lance05a,lorenz06a}. Therefore, we choose the probability distribution $P(y|x)$ to be parameterized as
\begin{equation}\label{gausscond}
P(y|x)=\frac{1}{\sqrt{2\pi V_{Y_{q}|x}^{2}}}\exp \left[\frac{-(\sqrt{\eta }\alpha
_{x}-y)^{2}}{2V_{Y_{q}|x}^{2}}\right]\;.
\end{equation}
Here, $\eta$ is the observed channel transmission, the amplitude $\alpha_0=-\alpha_1$ is chosen to be real. In this parameterization, the value of $\kappa$ as defined in Eq. (\ref{defkappa}) is given by by the loss of the quantum channel and the overlap of Alice's input states as
\begin{equation}
  \kappa=\left|\langle \sqrt{\eta}\alpha|-\sqrt{\eta}\alpha\rangle\right|\;.
\end{equation}
Furthermore, we assume that Bob observes the same variance (\ref{quadvar}) in his measured data for both the $q$- and the $p$- quadratures, so that 
\begin{equation}
V_{Y_{q}|x}^{2}=V_{Y_{p}|x}^{2}\;.
\end{equation} 
Here, we use the convention for the excess noise $\delta$ given in Ref. \cite{namiki04a}:
\begin{equation}
\delta=\frac{V_{Y_{q}|x}^{2}}{V_{Y_{q}|x, \mathrm{Vac}}^{2}}-1
\end{equation}
The quantity ${V_{Y_{q}|x,\mathrm{Vac}}^{2}}=\frac{1}{2}$ is the quadrature variance of the vacuum state. As the \emph{a priori} probabilities $p(x)=\frac{1}{2}$ are fixed, the probability distribution $p(y)$ is can easily be evaluated with the help of (\ref{gausscond}) and the secret key rate can be evaluated according to (\ref{final key rate}). Fig. (\ref{figkey}) shows our numerical results for the secret key rate versus the loss $1-\eta$ and different values for the excess noise $\delta$ in this typical scenario.

For the simulation, we assume that Bob announces the modulus of his measurement outcomes $y$ as $u=|y|$. The values of $\tilde{y}$ are determined by the map $\tilde{y}=0$ if $y<0$ and $\tilde{y}=1$ otherwise. After the announcement, the conditional mutual information between Alice and Bob is
\begin{align}
  I(X:\tilde{Y}|U)&=H(X|U)+H(X|\tilde{Y}U)\\&=H(X)-H(X|Y)\nonumber\\&=I(X:Y)\;.\nonumber
\end{align}
The announcement  $u=|y|$ contains no information about the bit-value $x$ for symmetric probability distributions like (\ref{gausscond}) as the conditional probability  $p(u|x)$ for a particular announcement $u$ is independent of $x$. Therefore it follows that $H(X|U)=H(X)$. The knowledge of Bob's measured outcome $y$ is obviously equivalent to the knowledge of $u=|y|$ and the sign of $y$, so that we have $H(X|\tilde{Y}U)=H(X|Y)$.
Therefore, we can achieve $I(X:\tilde{Y}|U)=I(X:Y)$ with this simple map as long as the probability distribution satisfies the symmetry condition $p(x|u)=1/2$.

For the numerical evaluation we optimize the secret key rate $G$ over the
overlap $\langle -\alpha |\alpha \rangle $ of the input states. In the
optimization we vary $\alpha $ between zero and 1 with step-width 0.05. For
each $\alpha$ we find the maximum of $s(\tilde{\varepsilon}_{x},\gamma)$ over all $\tilde{\varepsilon}_{x}\leq \varepsilon _{x}\leq U_{x}$ and $%
d_{l}\leq \gamma \leq d_{u}$. We find numerically that the maximum of $s(\tilde{\varepsilon}_{x},\gamma)$ is attained at the point $\gamma =d_{l}$.

Fig. (\ref{figkey}) shows the results of our simulation. As we can see, the
secret key rate is very susceptible to noise, whereas it coincides
with the optimal bound given in Ref. \cite{heid06a} for lossy but
noiseless quantum channels. However, one should keep in mind that
we only calculated an upper bound for Eve's knowledge, which we
expect not to be tight for finite excess noise. We have
bounded all three terms in Eq. (\ref{SYE}) separately rather than
bounding those terms simultaneously. Furthermore, one might expect
to find a different purification for the system $Q$ to make the
bound (\ref{SXE}) tighter. Finally, we have linearized the
conditional entropy $S(E|Y)$ in Section III. B in order to be able to find a bound.
However, the error introduced here might be quite large. 

 \section{Conclusion}

We have evaluated a lower bound to the secret key rate for a binary modulated CV-QKD protocol in the collective attack scenario. The analysis can be applied to any given channel noise, as Alice and Bob can estimate the conditional probability distribution $p(y|x)$ of their measurement outcomes arbitrary well in the limit that the number of exchanged signals tends to infinity. For any given probability distribution, the secret key rate can be computed according to Eq. (\ref{final key rate}). Although we demonstrate that our approach yields positive secret key rates for the case of small Gaussian excess noise, these results are not satisfying from a practical point of view, as the secret key rates drop quickly with increasing excess noise. Typically, the dominant contribution to the excess noise in CV-QKD experiments originate from noisy detectors. Our numerical results therefore indicate that it is necessary to analyze these kind of schemes in a trusted device scenario, if one wants to drop the assumption of ideal detectors and obtain secret rates of practical relevance. In this scenario, Eve cannot exploit the noise added by the detectors.

There are several options to make the protocol more robust against
channel excess noise. One could use more input states in order to test the quantum channel between Alice and Bob more efficiently and consequently limit Eve's possible interaction with the signal states. If one compares the secret key rates of Fig. (\ref{figkey})
with those given in Ref. \cite{heid07a} which correspond to a protocol using a Gaussian modulated, continuous set of input states
and a quantum channel imposing Gaussian noise onto the signal states, one realizes that the robustness of the secret key rate
increases by orders of magnitude. An introduction of a postselection step in the protocol can help to increase the
performance as well.

The authors want to thank M. Christandl,  M. Razavi, H. H\"aseler, T. Moroder and G. O. Myhr for helpful discussions. Y.-B. Zhao especially wants to thank Z.-F. Han and G.-C. Guo for supporting his visit to the Institute of Quantum Computing and many fruitful discussions on this topic.

This work was supported by the National Fundamental Research Program of China under Grant No 2006CB921900, the National Natural Science Foundation of China under Grants No. 60537020 and 60621064, the Knowledge Innovation Project of the Chinese Academy of Sciences (CAS), the European Union through the IST Integrated Project SECOQC, the NSERC Innovation Platform Quantum Works, the NSERC Discovery Grant and the Spanish Research Directorate, Grant FIS2005-06714.

\appendix

\section{Cauchy-Schwarz-Buniakowsky inequality}

\label{extreme}

The Cauchy-Schwarz-Buniakowsky inequality states \cite{gradshteyn94a} that for any two integrable functions $f(x)$ and $g(x)$
\begin{equation}\label{cauchy}
\left(\int_{a}^{b} dy f(y)h(y)\right)^{2}\leq\left(\int_{a}^{b} dy f^{2}(y)\right)\left(\int_{a}^{b} dy h^{2}(y)\right)
\end{equation}
holds. Application of inequality (\ref{cauchy}) to the left hand side of expression (\ref{blubber}) yields
\begin{eqnarray}
&&\int dy P(y)g[P(0|y),\gamma ]\sqrt{\tilde{\varepsilon}_{x}}b_{y}^{x}  \notag
\\
&=&\sqrt{\tilde{\varepsilon}_{x}}\int dy\underset{f(y)}{\underbrace{\sqrt{%
P(y|x)}b_{y}^{x}}}\underset{h(y)}{\underbrace{\{P(y) g[P(0|y),\gamma ]/\sqrt{%
P(y|x)}\}}}  \notag \\
&\leq &\sqrt{\tilde{\varepsilon}_{x}}\sqrt{\int dy P(y)g^{2}[P(0|y),\gamma ]%
\frac{P(y)}{P(y|x)}}.  \label{whatever}
\end{eqnarray}
Since one can rewrite the conditional probability $P(y|x)$ as $%
P(y|x)=P(x|y)P(y)/P(x)$ by using Bayes' rule and the \emph{a priori}
probabilities are given by $P(x)=\frac{1}{2}$, we have
\begin{equation}\label{bayes}
\frac{P(y)}{P(y|x)}=\frac{1}{2P(x|y)}\;
\end{equation}
and inequality (\ref{blubber}) follows from Eq. (\ref{whatever}) and Eq. (\ref{bayes}).

Similarly, one can evaluate the left hand side of Eq. (\ref{e10}) with the condition (\ref{e4}) as
\begin{eqnarray}  \label{a1}
&&\int dy P(y)g\left[P(0|y)\gamma\right]\sqrt{\tilde{\varepsilon}_{0}\tilde{\varepsilon}_{1}}b_{y}^{0}b_{y}^{1} \\
&=&\sqrt{\tilde{\varepsilon}_{0}}\sqrt{\tilde{\varepsilon}_{1}}\int dy\underset{f(y)}{\underbrace{\sqrt{P(y|1)}b_{y}^{1}}}\underset{h(y)}{\underbrace{\{P(y) g[P(0|y),\gamma ]/\sqrt{P(y|1)}\}}}  \notag \\
&\leq &\frac{\sqrt{\tilde{\varepsilon}_{0}\tilde{\varepsilon}_{1}}}{2}\sqrt{\int dy P(y|0){b_{y}^{0}}^{2}\frac{g^{2}\left[P(0|y),\gamma\right]}{P(0|y)P(1|y)}}\;,
\notag
\end{eqnarray}
where we used Eq. (\ref{bayes}) again in the last step.
Furthermore one can show that
\begin{align}  \label{a2}
&\int dy P(y|0){b_{y}^{0}}^{2}\frac{g^2\left[P(0|y),\gamma\right]}{P(0|y)P(1|y)}
\notag \\
&\leq \max_{y}\left\{\frac{g^{2}\left[P(0|y),\gamma\right]}{P(0|y)P(1|y)}\right\} \\
&=4g^{2}\left[\frac{1}{2},\gamma\right] \;.  \notag
\end{align}
The first line of Eq. (\ref{a2}) again follows from the boundary condition (\ref{e4}) for any integrable and bounded function $\frac{g^{2}_{y}(\gamma)}{P(0|y)P(1|y)}$. The second step can be shown by an
involved but straight forward calculation. Eq. (\ref{e10}) now follows from
Eqs. (\ref{a1}) and (\ref{a2}).

\section{Estimation of the mixedness $\varepsilon_{x}$ via homodyne measurements}\label{mix}

In this Appendix, we prove that the parameter $\varepsilon_{x}$ as defined in Eq. (\ref{fidelity}) can be estimated via Bob's homodyne measurements as
\begin{align}\label{esteps}
  \varepsilon_{x} &\leq \frac{1}{2}\left[\left(V^{2}_{y_{q}|x}+\frac{1}{2}\right)\left(V^{2}_{y_{p}|x}+\frac{1}{2}\right)-1\right]=\frac{1}{2}(W-1)\;.
\end{align}
The mixedness $\varepsilon_{x}$ is given by the fidelity between Bob's received state $\rho_{B}^{x}$ and the pure coherent state $\overline{\beta_{x}}$ as
\begin{equation}\label{lofid}
  \langle\overline{\beta}_{x}|\rho_{B}^{x}|\overline{\beta}_{x}\rangle = 1-\varepsilon_{x}\;.
\end{equation}
The amplitude $\overline{\beta}_{x}$ is given by Eq. (\ref{defbetabar}) and we use the convention (\ref{quad}) for quadrature operators.  The conditional variances $V^{2}_{y_{q}|x}$ and $V^{2}_{y_{p}|x}$ are then given by
\begin{align}\label{condvar}
  V^{2}_{y_{q}|x}&=\mathrm{tr}\left(\rho_{B}^{x}\hat{q}^{2}\right)-\left[\mathrm{tr}\left(\rho_{B}^{x}\hat{q}\right)\right]^{2}\\
  V^{2}_{y_{p}|x}&=\mathrm{tr}\left(\rho_{B}^{x}\hat{p}^{2}\right)-\left[\mathrm{tr}\left(\rho_{B}^{x}\hat{p}\right)\right]^{2}\nonumber
\end{align}

Let us introduce a state $\overline{\rho}=\hat{D}(-\overline{\beta}_{x})\rho_{B}^{x}\hat{D}(\overline{\beta}_{x})$ with zero mean values for the quadrature operators (\ref{quad}) to simplify the analysis. Here $\hat{D}(\overline{\beta}_{x})$ denotes the displacement operator according to the amplitude $\overline{\beta}_{x}$. Obviously,
\begin{equation}\label{fid2}
  \langle\overline{\beta}|\rho_{B}^{x}|\overline{\beta}\rangle=\langle 0|\overline{\rho}|0\rangle=1-\varepsilon_{x}\;
\end{equation}
holds, whereas $|0\rangle$ denotes the vacuum state. The variances (\ref{condvar}) can now be evaluated with the definition (\ref{quad}) as
\begin{align}\label{condvar2}
  V^{2}_{y_{q}|x}&=\mathrm{tr}\left(\overline{\rho}\hat{q}^{2}\right)=\frac{1}{2}\mathrm{tr}\left[\overline{\rho}\left(2\hat{n}+1+\hat{a}^{2}+\hat{a}^{\dagger 2}\right)\right]\\
  V^{2}_{y_{p}|x}&=\mathrm{tr}\left(\overline{\rho}\hat{p}^{2}\right)\nonumber=\frac{1}{2}\mathrm{tr}\left[\overline{\rho}\left(2\hat{n}+1-\hat{a}^{2}-\hat{a}^{\dagger 2}\right)\right]\nonumber\;,
\end{align}
where we have introduced the photon number operator $\hat{n}=\hat{a}^{\dagger}\hat{a}$ as short hand notation. The quantity $W$ in Eq. (\ref{esteps}) now reads
\begin{align}\label{dabblju}
W=&\frac{1}{4}\mathrm{tr}\left[\overline{\rho}\left(2\hat{n}+2+\hat{a}^{2}+\left(\hat{a}^{\dagger}\right)^{2}\right)\right]\\
  &\times\mathrm{tr}\left[\overline\rho\left(2\hat{n}+2-\hat{a}^{2}-\left(\hat{a}^{\dagger}\right)^{2}\right)\right]\nonumber\\
  =&\left[\mathrm{tr}\left(\overline{\rho}\hat{n}\right)+1\right]^2-\frac{1}{4}\mathrm{tr}\left[\overline{\rho}\left(\hat{a}^{2}+\hat{a}^{\dagger 2}\right)\right]^{2}\nonumber\;.
\end{align}
We proceed in rewriting the last term in (\ref{dabblju}) with the help of Eqs. (\ref{condvar2}) in the Fock-basis $\{|n\rangle\}_{n}$ as
\begin{align}\label{a2a+2}
  \mathrm{tr}\left[\overline{\rho}\left(\hat{a}^{2}+\hat{a}^{\dagger 2}\right)\right]=&\sum_{n=0}^{\infty}\sqrt{n+2}\sqrt{n+1} \langle n+2|\overline{\rho}|n\rangle\\
&+\sum_{n=2}^{\infty}\sqrt{n}\sqrt{n-1} \langle n-2|\overline{\rho}|n\rangle\nonumber\\
=&2\sum_{n=0}^{\infty}\sqrt{n+2}\sqrt{n+1} \mathrm{Re}\langle n+2|\overline{\rho}|n\rangle\nonumber\;.
\end{align}
Since $\langle i|\overline{\rho}|j\rangle$ is a positive semidefinite matrix, any principal minor is a positive semidefinite matrix. It follows that
\begin{equation}\label{prinmin}
  \langle i|\overline{\rho}|i\rangle \langle j|\overline{\rho}|j\rangle- \left|\langle i|\overline{\rho}|j\rangle\right|^{2} \geq 0\;,
\end{equation}
as this can be interpreted as the determinant of the 2 by 2 principal minor that arises by only keeping the $i$-th and $j$-th entries. The positivity of this determinant then follows by realizing that the determinant is just the product of the non-negative eigenvalues of the corresponding principal minor \cite{horn85a}. The result (\ref{prinmin}), together with the triangle inequality, can be used to estimate the modulus of Eq. (\ref{a2a+2}) as
\begin{align}\label{moda2a+2}
\left|\mathrm{tr}\left[\overline{\rho}\left(\hat{a}^{2}+\hat{a}^{\dagger 2}\right)\right]\right|\leq& 2\sum_{n=0}^{\infty}\sqrt{n+2}\sqrt{n+1}\sqrt{\langle n|\overline{\rho}|n\rangle}\\
&\times \sqrt{\langle n+2|\overline{\rho}|n+2\rangle}\nonumber\\
\leq& 2\sqrt{\sum_{n=0}^{\infty}(n+1)\langle n|\overline{\rho}|n\rangle}\nonumber\\
&\times \sqrt{\sum_{n=0}^{\infty}(n+2)\langle n+2|\overline{\rho}|n+2\rangle}\nonumber\\
=& 2 \sqrt{\mathrm{tr}\left(\overline{\rho}\hat{n}\right)+1}\sqrt{\mathrm{tr}\left(\overline{\rho}\hat{n}\right)-\langle 1|\overline{\rho}|1\rangle}\;.
\end{align}
The second estimation in Eq. (\ref{moda2a+2}) follows from the Cauchy-Schwarz inequality. Inserting this result into Eq. (\ref{dabblju}) yields
\begin{align}\label{dabblju2}
  W \geq \left[\mathrm{tr}\left(\overline{\rho}\hat{n}\right)+1\right]\left[1+\langle1|\overline{\rho}|1\rangle\right]\;.
\end{align}
We need to find a lower bound on Eq. (\ref{dabblju2}) depending only on $\varepsilon_{x}$. Note that $\mathrm{tr}\left(\overline{\rho} \hat{n}\right)$ can be written as
\begin{align}
  \mathrm{tr}\left(\overline{\rho}\hat{n}\right)=&\langle{1}|\overline{\rho}|1\rangle+\sum_{n=2}^{\infty}\langle{n}|\overline{\rho}|n\rangle n\\
 \geq& \langle{1}|\overline{\rho}|1\rangle +2\left( \varepsilon_{x} -\langle{1}|\overline{\rho}|1\rangle\right)\;.\nonumber
\end{align}
As it can be seen from Eq. (\ref{fid2}), the fidelity of $\overline{\rho}$ with the vacuum is $1-\varepsilon_{x}$. It follows that all matrix elements $\langle n|\overline{\rho}|n \rangle$ for $n\geq 1$ sum up to $\varepsilon_{x}$, so that $\sum_{n=2}^{\infty}\langle{n}|\overline{\rho}|n\rangle n$ is minimal if all $\langle n|\overline{\rho}|n\rangle=0$ except for $\langle 2|\overline{\rho}|2\rangle$, which has then to be equal to $\varepsilon_{x}-\langle 1|\overline{\rho}|1\rangle$ by the summing condition. Therefore, Eq. (\ref{dabblju2}) can be estimated as
\begin{align}\label{dabblju3}
  W&\geq \left(1+2\varepsilon_{x}-\langle 1|\overline{\rho}|1\rangle\right)\left(1+\langle 1|\overline{\rho}|1\rangle\right)\\
&=1+2\varepsilon_{x}+\langle 1|\overline{\rho}|1\rangle(2\varepsilon_{x}-\langle 1|\overline{\rho}|1\rangle)\nonumber\;,
\end{align}
As $0 \leq \langle 1|\overline{\rho}|1\rangle\leq \varepsilon_{x}$, the last term of Eq. (\ref{dabblju3}) is never negative and equal to zero iff  $\langle 1|\overline{\rho}|1\rangle= 0$. It follows that
\begin{align}
  W\geq 1+2\varepsilon_{x}\;.
\end{align}
Inserting this result into Eq. (\ref{esteps}) concludes the proof.

\section{Estimation to the overlap of Bob's maximal eigenstates}\label{estovereigen}

In the following, we derive explicit expressions for the bounds to the overlap $|\langle\tilde{\beta}_{0}|\tilde{\beta}_{1}\rangle$ of Bob's conditional eigenstates to the maximal eigenvalue $1-\tilde{\varepsilon}_{x}$ as given by expression (\ref{e17}). Assume that we know the fidelity
\begin{equation}
  \langle \overline{\beta}_{x}|\rho_{B}^{x}|\overline{\beta}_{x}\rangle = 1-\varepsilon_{x}
\end{equation}
of Bob's received state $\rho_{B}^{x}$ with the coherent state $|\overline{\beta_{x}}\rangle$ is given. The amplitude $\overline{\beta_{x}}$ is defined in Eq. (\ref{defbetabar}).
We can express the conditional states $\rho_{B}^{x}$ in a natural basis of displaced Fock-states $\{|\phi^{x}_{k}\rangle\}=\{D(\overline{\beta}_{x})|k\rangle\}$. Here, the the parameter $k$ labels the photon number. Obviously, $|\overline{\beta}_{x}\rangle=|\phi^{x}_{0}\rangle$ holds. In this basis, $\rho_{B}^{x}$ reads
\begin{equation}\label{reprbobnat}
\rho_{B}^{x}=\left[
\begin{array}{ccc}
a_{00}&a_{01}&... \\
a_{01}^{*}& a_{11}&\\
\vdots&&\ddots
\end{array}
\right]= V^{x}D^{x}{V^{x}}^{\dagger}\;,
\end{equation}
where $V^{x}$ denotes a unknown unitary matrix and $D^{x}$ is the representation of $\rho_{B}^{x}$ in its eigenbasis. Without loss of generality, we can choose the first element in the $D$-Matrix to correspond to the biggest eigenvalue, so that 
\begin{equation}\label{D00}
D^{x}_{00}=1-\tilde{\varepsilon}_{x}\;.
\end{equation} 
From Eq. (\ref{reprbobnat}), we know that
\begin{align}\label{a00}
1-\varepsilon_{x}= a_{00}&=|V_{00}^{x}|^{2}D^{x}_{00}+\sum_{k=1}^{\infty}|V_{0k}^{x}|^2 D^{x}_{kk}
\end{align}
As $V^{x}$ is unitary, it follows that 
\begin{equation}
\sum_{k=1}^{\infty}|V_{0k}^{x}|^2=1-\left|V_{00}^{x}\right|^{2}\;.
\end{equation}
Moreover, $D^{x}$ is normalized, so that
\begin{equation}
 \sum_{k=1}^{\infty} D^{x}_{kk}=1-D^{x}_{00}=\tilde{\varepsilon}_{x}\;,
\end{equation}
where we used Eq. (\ref{D00}). This can be used to bound the infinite sum in Eq. (\ref{a00}) as
\begin{equation}\label{maaan}
  \sum_{k=1}^{\infty}|V_{0k}^{x}|^2 D^{x}_{kk}\leq \left(1-\left|V_{00}^{x}\right|^{2}\right)\tilde{\varepsilon}_{x}\;,
\end{equation}
since all terms $|V_{0k}^{x}|^2$ and $D^{x}_{kk}$ appearing in the sum are strictly positive. Therefore, we can bound Eq. (\ref{a00}) according to inequality (\ref{maaan}) as
\begin{align}
1-\varepsilon_{x}&\leq \left|V_{00}^{x}\right|^{2}(1-\tilde{\varepsilon}_{x})+\left(1-\left|V_{00}^{x}\right|^{2}\right)\tilde{\varepsilon}_{x}\nonumber\\
&=\left|V_{00}^{x}\right|^{2}(1-2\tilde{\varepsilon}_{x})+\tilde{\varepsilon}_{x}\nonumber\;.
\end{align}
It follows that
\begin{equation}\label{lowV}
  \left|V_{00}^{x}\right|^{2}\geq\frac{1-\varepsilon_{x}-\tilde{\varepsilon}_{x}}{1-2\tilde{\varepsilon}_{x}}\;.
\end{equation}
Moreover, one can use Eq. (\ref{a00}) to obtain a lower bound on $\left|V_{00}^{x}\right|^{2}$ as
\begin{equation}
  1-\varepsilon_{x}\geq \left|V_{00}^{x}\right|^{2}(1-\tilde{\varepsilon}_{x})\;,
\end{equation}
so that
\begin{equation}\label{upV}
  \left|V^{x}_{00}\right|^{2}\leq\frac{1-\varepsilon_{x}}{1-\tilde{\varepsilon}_{x}}\;.
\end{equation}
On the other hand, Bob's conditional states can be written as
\begin{align}\label{exp1}
  \rho_{B}^{0}&=V^{0}D^{0}{V^{0}}^{\dagger}\\
  \rho_{B}^{1}&=U V^{1}D^{1}{V^{1}}^{\dagger} U^{\dagger}\nonumber\;,
\end{align}
where the unitary operation $U$ is given, up to an unimportant unimodular phase, by $U=\hat{D}\left(\overline{\beta}_{1}\right)\hat{D}\left(-\overline{\beta}_{0}\right)$ and $\hat{D}$ denotes the displacement operator. Let us denote the eigenvectors of Bob's conditional states $\rho_{B}^{x}$ as $\{|\tilde{\beta}_{l}^{x}\rangle\}$ with $|\tilde{\beta}_{x}\rangle$ being the eigenstate corresponding to the biggest eigenvalue $1-\tilde{\varepsilon}_{x}$. With the representation (\ref{exp1}), these states can be written as
\begin{align}
  |\tilde{\beta}_{0}\rangle&=V^{0}|\phi_{0}^{0}\rangle\\
  |\tilde{\beta}_{1}\rangle&=U V^{1}|\phi_{0}^{0}\rangle\nonumber\;,
\end{align}
so that
\begin{align}
  \langle \tilde{\beta}_{0}|\tilde{\beta}_{1}\rangle=\sum_{k,j=0}^{\infty}{V^{0}_{k0}}^{*}U_{kl}V_{l0}^{1}
\end{align}
By  use of the triangle inequalities, one can construct an upper bound as
\begin{align}\label{CSu}
  \left| \langle\tilde{\beta}_{0}|\tilde{\beta}_{1}\rangle\right|\leq& \left|U_{00}\right|\left|V_{00}^{0}\right|\left|V_{00}^{1}\right|+\left|V_{00}^{0}\right|\left|\sum_{l=1}^{\infty} U_{0l}V_{l0}^{1}\right|\\
&+\left|V_{00}^{1}\right|\left|\sum_{k=1}^{\infty} U_{k0}{V_{k0}^{0}}^{*}\right|+\left|\sum_{k,l=1}^{\infty} {V_{k0}^{0}}^{*}U_{kl}{V_{l0}^{1}}\right|\nonumber
\end{align}
and a lower bound as
\begin{align}\label{CSl}
  \left| \langle\tilde{\beta}_{0}|\tilde{\beta}_{1}\rangle\right|\geq& \left|U_{00}\right|\left|V_{00}^{0}\right|\left|V_{00}^{1}\right|-\left|V_{00}^{0}\right|\left|\sum_{l=1}^{\infty} U_{0l}V_{l0}^{1}\right|\\
&-\left|V_{00}^{1}\right|\left|\sum_{k=1}^{\infty} U_{k0}{V_{k0}^{0}}^{*}\right|-\left|\sum_{k,l=1}^{\infty} {V_{k0}^{0}}^{*}U_{kl}{V_{l0}^{1}}\right|\nonumber\;.
\end{align}
Upper bounds on the sums in Eqs.(\ref{CSu}) and (\ref{CSl}) can be obtained by using the Cauchy-Schwarz inequality as
\begin{align}\label{CS2}
  \sum_{k=1}^{\infty}\left|U_{k0}\right|\left|V_{k0}^{0}\right|&\leq \sqrt{1-\left|U_{00}\right|^{2}}\sqrt{1-\left|V_{00}^{0}\right|^{2}}\\
  \sum_{l=1}^{\infty}\left|U_{0l}\right|\left|V_{l0}^{1}\right|&\leq \sqrt{1-\left|U_{00}\right|^{2}}\sqrt{1-\left|V_{00}^{1}\right|^{2}}\nonumber\\
\left|\sum_{k,l=1}^{\infty} {V_{k0}^{0}}^{*}U_{kl}{V_{l0}^{1}}\right|&\leq \sqrt{1-\left|V_{00}^{0}\right|^{2}}\sqrt{1-\left|V_{00}^{1}\right|^{2}}\nonumber\;.
\end{align}
It is easy to see that $|U_{00}|=|\langle \overline{\beta}_{0}|\overline{\beta}_{1}\rangle|:=\kappa$.
Finally, inserting Eqs. (\ref{lowV}, \ref{upV}, \ref{CS2}) in Eq. (\ref{CSu}) yields
\begin{align}\label{cup}
  \left|\langle\tilde{\beta}_{0}|\tilde{\beta}_{1}\rangle\right|\leq& \kappa \sqrt{\frac{1-\varepsilon_{0}}{1-\tilde{\varepsilon}_{0}}}\sqrt{\frac{1-\varepsilon_{1}}{1-\tilde{\varepsilon}_{1}}}\\
&+\sqrt{1-\kappa^{2}}\sqrt{\frac{1-\varepsilon_{0}}{1-\tilde{\varepsilon}_{0}}}\sqrt{\frac{\varepsilon_{1}-\tilde{\varepsilon}_{1}}{1-2\tilde{\varepsilon}_{1}}}\nonumber\\
&+\sqrt{1-\kappa^{2}}\sqrt{\frac{1-\varepsilon_{1}}{1-\tilde{\varepsilon}_{1}}}\sqrt{\frac{\varepsilon_{0}-\tilde{\varepsilon}_{0}}{1-2\tilde{\varepsilon}_{0}}}\nonumber\\
&+\sqrt{\frac{\varepsilon_{1}-\tilde{\varepsilon}_{1}}{1-2\tilde{\varepsilon}_{1}}}\sqrt{\frac{\varepsilon_{0}-\tilde{\varepsilon}_{0}}{1-2\tilde{\varepsilon}_{0}}}\nonumber\;.
\end{align}
Similarly, a lower bound can be obtained by Eqs. (\ref{lowV}, \ref{upV}, \ref{CS2}) and (\ref{CSl}) as
\begin{align}\label{clow}
  \left|\langle\tilde{\beta}_{0}|\tilde{\beta}_{1}\rangle\right|\geq& \kappa \sqrt{\frac{1-\varepsilon_{0}-\tilde{\varepsilon}_{0}}{1-2\tilde{\varepsilon}_{0}}}\sqrt{\frac{1-\varepsilon_{1}-\tilde{\varepsilon}_{1}}{1-2\tilde{\varepsilon}_{1}}}\\
&-\sqrt{1-\kappa^{2}}\sqrt{\frac{1-\varepsilon_{0}}{1-\tilde{\varepsilon}_{0}}}
\sqrt{\frac{\varepsilon_{1}-\tilde{\varepsilon}_{1}}{1-2\tilde{\varepsilon}_{1}}}\nonumber\\
&-\sqrt{1-\kappa^{2}}\sqrt{\frac{1-\varepsilon_{1}}{1-\tilde{\varepsilon}_{1}}}\sqrt{\frac{\varepsilon_{0}-\tilde{\varepsilon}_{0}}{1-2\tilde{\varepsilon}_{0}}}\nonumber\\
&-\sqrt{\frac{\varepsilon_{1}-\tilde{\varepsilon}_{1}}{1-2\tilde{\varepsilon}_{1}}}\sqrt{\frac{\varepsilon_{0}-\tilde{\varepsilon}_{0}}{1-2\tilde{\varepsilon}_{0}}}\nonumber\;.
\end{align}
The explicit expression for $c_{l}\left(\tilde{\varepsilon}_{x},\varepsilon_{x}, \kappa\right)$ is therefore given by Eq. (\ref{clow}) and respectively, $c_{u}\left(\tilde{\varepsilon}_{x},\varepsilon_{x}, \kappa\right)$ is given by Eq. (\ref{cup}).

\section{Estimation to the overlap of Eve's maximal eigenstates}

\label{estimation}

In the collective attack scenario, Eve's attack can be modelled by attaching an ancilla system to the signals $|\pm \alpha \rangle$ and performing a unitary operation on the joint system. As any unitary preserves the inner product, the overlap $|\langle \Psi _{BE}^{0}|\Psi_{BE}^{1}\rangle|$ of the states after the interaction is given by input overlap $|\langle -\alpha |\alpha \rangle |$. This can be written as
\begin{eqnarray}  \label{innerprod}
|\langle -\alpha |\alpha \rangle | &=&|\langle \Psi _{BE}^{0}|\Psi
_{BE}^{1}\rangle| \\
&=&|\sqrt{(1-\tilde{\varepsilon}_{0})(1-\tilde{\varepsilon}_{1})}\langle
\tilde{\beta}_{0}|\tilde{\beta}_{1}\rangle \langle \tilde{\varepsilon}_{0}|\tilde{\varepsilon}_{1}\rangle  \notag \\
&&+\sqrt{(1-\tilde{\varepsilon}_{0})\tilde{\varepsilon}_{1}}\langle \tilde{\beta}_{0}|\langle \tilde{\varepsilon}_{0}|\varphi _{EB}^{1}\rangle  \notag \\
&&+\sqrt{(1-\tilde{\varepsilon}_{1})\tilde{\varepsilon}_{0}}\langle \varphi
_{EB}^{0}|\tilde{\beta}_{1}\rangle |\tilde{\varepsilon}_{1}\rangle  \notag \\
&&+\sqrt{\tilde{\varepsilon}_{1}\tilde{\varepsilon}_{0}}\langle \varphi
_{EB}^{0}|\varphi _{EB}^{1}\rangle |.  \notag
\end{eqnarray}%
using decomposition (\ref{stateBE}), where $|\varphi _{EB}^{x}\rangle $ is orthogonal to $|\tilde{\beta}_{x}\rangle |\tilde{\varepsilon}_{x}\rangle$. This orthogonality can be used to construct the inequalities
\begin{align}  \label{propover}
|\langle\varphi _{EB}^{0}|\varphi _{EB}^{1}\rangle |^{2}+|\langle
\varphi_{EB}^{0}|\tilde{\beta}_{1}\rangle |\tilde{\varepsilon}_{1}\rangle |^{2}&\leq 1 \\
|\langle \varphi _{EB}^{0}|\varphi _{EB}^{1}\rangle |^{2}+|\langle
\tilde{\beta}_{0}|\langle \tilde{\varepsilon}_{0}|\varphi _{EB}^{1}\rangle |^{2}&\leq 1\;.
\notag
\end{align}
We can estimate the last three terms of the right hand side of Eq. (\ref{innerprod}) using the triangle inequality and inequalities (\ref{propover}) as
\begin{align}  \label{b1}
|&\sqrt{(1-\tilde{\varepsilon}_{0})\tilde{\varepsilon}_{1}}\langle
\tilde{\beta}_{0}|\langle \tilde{\varepsilon}_{0}|\varphi _{EB}^{1}\rangle \\
&+\sqrt{(1-\tilde{\varepsilon}_{1})\tilde{\varepsilon}_{0}}\langle
\varphi_{EB}^{0}|\tilde{\beta}_{1}\rangle |\tilde{\varepsilon}_{1}\rangle+  \notag \\
&+\sqrt{\tilde{\varepsilon}_{1}\tilde{\varepsilon}_{0}}\langle \varphi
_{EB}^{0}|\varphi_{EB}^{1}\rangle |  \notag \\
\leq& \underset{x_{0}}{\underbrace{\sqrt{1-|\langle \varphi
_{EB}^{0}|\varphi_{EB}^{1}\rangle |^{2}}}}\underset{y_{0}}{\underbrace{\left(%
\sqrt{(1-\tilde{\varepsilon}_{0})\tilde{\varepsilon}_{1}}+\sqrt{(1-\tilde{%
\varepsilon}_{1})\tilde{\varepsilon}_{0}}\right)}}  \notag \\
&+\underset{x_{1}}{\underbrace{|\langle \varphi
_{EB}^{0}|\varphi_{EB}^{1}\rangle |}}\underset{y_{1}}{\underbrace{\sqrt{%
\tilde{\varepsilon}_{1}\tilde{\varepsilon}_{0}}}}  \notag \\
\leq& \sqrt{\lbrack \sqrt{(1-\tilde{\varepsilon}_{1})\tilde{\varepsilon}_{0}}%
+\sqrt{(1-\tilde{\varepsilon}_{0})\tilde{\varepsilon}_{1}}]^{2}+\tilde{%
\varepsilon}_{1}\tilde{\varepsilon}_{0}}\;.  \notag
\end{align}%
In the sixth line we have use fact the that if $\sum_{i}x_{i}^{2}=1$, $%
\sum_{i}x_{i}y_{i}\leq \sqrt{\sum_{i}y_{i}^{2}}$ holds, which can easily derived from the Cauchy-Schwarz inequality of two vectors in $\mathbbm{R}^{2}$. From Eq. (\ref{b1})
and Eq. (\ref{innerprod}), we obtain
\begin{eqnarray*}
&&|\langle \tilde{\beta}_{0}|\tilde{\beta}_{1}\rangle \langle \tilde{\varepsilon}_{0}|\tilde{\varepsilon}_{1}\rangle | \\
&\geq &\frac{|\langle -\alpha |\alpha \rangle |-\sqrt{[\sqrt{(1-\tilde{\varepsilon}_{1})\tilde{\varepsilon}_{0}}+\sqrt{(1-\tilde{\varepsilon}_{0})\tilde{\varepsilon}_{1}}]^{2}+\tilde{\varepsilon}_{1}\tilde{\varepsilon}_{0}}
}{\sqrt{(1-\tilde{\varepsilon}_{0})(1-\tilde{\varepsilon}_{1})}}.
\end{eqnarray*}
and
\begin{eqnarray*}
&&|\langle \tilde{\beta}_{0}|\tilde{\beta}_{1}\rangle \langle \tilde{\varepsilon}_{0}|\tilde{\varepsilon}_{1}\rangle | \\
&\leq &\frac{|\langle -\alpha |\alpha \rangle |+\sqrt{[\sqrt{(1-\tilde{%
\varepsilon}_{1})\tilde{\varepsilon}_{0}}+\sqrt{(1-\tilde{\varepsilon}_{0})%
\tilde{\varepsilon}_{1}}]^{2}+\tilde{\varepsilon}_{1}\tilde{\varepsilon}_{0}}%
}{\sqrt{(1-\tilde{\varepsilon}_{0})(1-\tilde{\varepsilon}_{1})}}.
\end{eqnarray*}%
Finally, we obtain Eqs. (\ref{e18}-\ref{du}) by inserting the extremal values for the possible overlaps $\left|\langle\tilde{\beta}_{0}|\tilde{\beta}_{1}\rangle\right|$ of Bob's maximal eigenstates given by Eq. (\ref{e17}) and the definition $\gamma :=|\langle \tilde{\varepsilon}_{0}|\tilde{\varepsilon}_{1}\rangle |$.

\end{document}